\documentclass[12pt,preprint]{aastex}

\shorttitle{X-ray Spectra of RS  Oph in Outburst}
\shortauthors{T. J. Nelson et al.}

\begin{document}
\title{X-ray Spectroscopy of the 2006 Outburst of RS  Oph}
\author{Thomas Nelson}
\affil{Department of Astronomy, University of Wisconsin-Madison, 475 N Charter St, 
Madison, WI, 53706, USA}
\email{nelson@astro.wisc.edu}
\author{Marina Orio}
\affil{INAF - Osservatorio Astronomico di Padova, vicolo Osservatorio, 5, I-35122 
Padova, Italy and Department of Astronomy, University of Wisconsin-Madison 475 N 
Charter St, Madison, WI, 53706, USA}
\author{Joseph P. Cassinelli}
\affil{Department of Astronomy, University of Wisconsin-Madison, 475 N Charter St, 
Madison, WI, 53706, USA}
\author{Martin Still}
\affil{South African Astronomical Observatory, PO Box 9, Observatory, Cape Town, 
South Africa}
\author{Elia Leibowitz}
\affil{Wise Observatory and the School of Physics and Astronomy, Tel Aviv University, 
Tel Aviv, 69978, Israel}
\author{Paola Mucciarelli}
\affil{Dipartimento di Astronomia, Universit\`a di Padova,
 vicolo Osservatorio, 2, I-35122 Padova, Italy}

\begin{abstract}
We present the X-ray grating spectra of the recurrent nova RS Ophiuchi during its 2006 outburst, obtained with the XMM-Newton and Chandra observatories.  Two weeks after optical maximum, the X-ray spectrum was hard and dominated by emission lines of H-like and He-like ions. The X-ray luminosity was 2.4 $\times$ 10$^{36}$ erg cm$^{-2}$ in the 0.33-10 keV range.  The spectra indicate a collisionally dominated plasma with a broad range of temperatures.  All the lines are blue-shifted, with the velocity shift increasing with lower ionization state and longer wavelength. Two weeks later, the spectrum was still dominated by emission lines, although the line ratios present indicate cooling.  During this observation, a soft X-ray flare occurred in which a new system of higher velocity emission lines appeared in the soft end of the spectrum. Towards the end of the second month the emission spectrum became composite, with both absorption and emission features. The dominant component during the third month was the supersoft continuum with the broad absorption features of a hot white dwarf atmosphere. The absorption lines did not appear to be significantly blue-shifted as previously observed in another classical nova.  A preliminary fit of an atmospheric model available in the literature indicates a white dwarf temperature slightly exceeding 800,000 K.  The X-ray luminosity reached at least 9 $\times$ 10$^{37}$ erg cm$^{-2}$ in the 0.2-1 keV range, while the intrinsic nebular absorption decreased by a factor of five since the first observation.  The spectral fits indicate a massive white dwarf, with a mass of at least 1.2 M$_{\odot}$.  Therefore, RS Oph may be an important type Ia supernova progenitor.  We show that the data are consistent with mass loss ending before day 54 after the outburst, and nuclear burning ending around day 69.  A rapid decay in X-ray luminosity  followed after week 10.  The X-ray luminosity 5, 7 and 8 months after optical maximum dropped by more than two orders of magnitude,  and the spectra are dominated by soft X-ray emission lines typical of a collisionally excited and ionized plasma.  They do not appear to consistent with emission from an accretion disk. 

\end{abstract}

\keywords{stars: individual (RS Ophiuchi), stars: novae, cataclysmic variables, X-rays: binaries, techniques: spectroscopic}

\section{Introduction}
RS Ophiuchi (RS Oph) is a member of a rare subclass of cataclysmic variable known 
as recurrent novae (RN).  As the name suggests, RN are systems which are observed 
to go into outburst more than once.  Nova outbursts are believed to be the result of 
thermonuclear runaway (TNR) on the surface of a white dwarf accreting material from 
a binary companion (e.g. Schatzman 1950, Starrfield et al. 1974, Prialnik et al. 1978).  
The TNR occurs once the pressure at the base of the shell of accreted material reaches a high enough value.  
Following Fujimoto (1982a) the critical pressure required for the TNR ignition is given by 
\begin{equation}
P_{c} = \frac{G M_{WD} \Delta M_{c}}{4 \pi R^{4}}
\end{equation} where M$_{WD}$ and R are the mass  and radius of the white dwarf, 
and $\Delta$M$_{c}$ is the mass of the accreted shell at the onset of the TNR.  As long as 
accretion continues, all novae are expected to recur with timescales ranging from 10$^
{6}$ years to only a few decades.  For systems hosting high mass white dwarfs, and 
those with higher accretion rates, critical envelope masses will be reached in a shorter 
time.  Therefore, the fact that we see the recurrent novae repeat at all 
during a human lifetime is an indication 
that we are dealing with higher mass white dwarfs.  If the white dwarfs in recurrent 
novae can grow in mass with each subsequent accretion--outburst cycle, they may 
eventually approach and even exceed the Chandrasekhar limit.  Therefore, RN are 
interesting and important supernova Ia progenitor candidates.

RS Oph went into its 6th recorded outburst on Feb 12.83, 2006 (Narumi 2006), having 
previously been detected in outburst in 1898, 1933, 1958, 1967 and 1985.  It is 
possible that other outbursts have occurred since the discovery of the system, but were 
missed by observers.  Long term lightcurves of RS Oph obtained by the AAVSO show 
that the system undergoes a characteristic dimming just before the outburst, and an 
overshoot at the end of decay from maximum beyond the mean inter-outburst magnitude (Oppenheimer 1993).  Schaefer (2004) 
reported that a reduction in magnitude was detected on plates obtained in 1907, just 
after the end of the seasonal observing gap.  He attributed these low values to the 
post-outburst dip observed in other events, and suggested that RS Oph experienced 
an outburst in 1907 while it was hidden by the sun.  Similarly, dips in magnitude were 
detected both before and after the 1945 observing gap, suggesting another outburst 
was missed at that time (Oppenheimer \& Mattei 1996).  

RS Oph is believed to host a massive white dwarf accreting material from a red giant 
companion of spectral type K7 (M\"urset \& Schmid 1999).  The most recent determination of
 the orbital 
period of the binary is 456 days (Fekel et al. 2000).  The secondary is believed to 
underfill its Roche Lobe, and so accretion proceeds through a wind and not through 
Roche Lobe Overflow (RLO), as in cataclysmic variables and classical novae (e.g. 
Dobrzycka et al. 1996, Hachisu et al. 2001).   The distance to the system has been 
discussed extensively in the literature, and disagreements remain.  Bode (1987) gives 
1.6 $\pm$ 0.3 kpc.   Results from the current outburst give a distance ranging from 540 
pc (Monnier et al. 2006) to 1.3--1.7 kpc (Hachisu et al. 2006b).  This large spread in 
distance makes the exact parameters of the system difficult to determine.   We will follow Bode in this work, and assume a distance of 
1.6 $\pm$ 0.3 kpc.

The system was observed in X-rays by EXOSAT on several occasions during the 1985 
outburst (Mason et al., 1987).  However, only low resolution spectra were obtained as 
X-ray grating spectroscopy was not available in 1985.  Since then the situation has 
vastly improved, with no less than 4 active X-ray observatories equipped with 
advanced instrumentation capable of high resolution grating spectroscopy or 
microsecond timing accuracy.  Data have been obtained at frequent intervals by the 
Rossi X-ray Timing Experiment (RXTE), the X-ray telescope (XRT) on board Swift, and 
of course Chandra and XMM-Newton.  In this paper, we present the Chandra and XMM-Newton data, 
which include the first grating spectra obtained of any of the recurrent novae.  The 
temporal coverage offered by such an extensive set of observations allows us for the 
first time to test models of the evolution of an outburst in the RS Oph system.  We 
present a detailed timing analysis of the data set in an accompanying paper (Leibowitz et 
al. 2007).  

\section{Observations and Data Reduction}
Thanks to a generous allocation of Director's Discretionary Time (DDT) , we were able to obtain a total of 9 grating observations of RS  Oph; 5 XMM-Newton and 4 Chandra pointings.  Table 1 outlines the details of each observations.  The quoted exposure times are the net exposures after being corrected for periods of bad telemetry. These observations were initially carried out every two weeks from days 14 to 67.  After that, one pointing each in June, September and October was obtained, following the evolution of the system back to its inter-outburst state.  The Chandra observations were carried out using the HETG/ACIS-S in February, and then the LETG/HRC-S in all subsequent pointings.  The data were reduced  using CIAO v. 3.3.1 with CALDB v. 3.2.2.  For each observation, the \textit{mkgrmf} task was used to create 
response matrices for each order, followed by \textit{mkgarf} to produce the ancillary 
response files.  We worked with the +1 and -1 orders for each grating, which contained most of the flux.  

The XMM-Newton observatory has a suite of 6 instruments; two Reflection Grating Spectrographs (RGS), 3 European Photon Imaging Cameras (EPIC), and the optical monitor (OM), an optical and near UV telescope.  The EPIC utlilize two slightly different technologies.  The EPIC-MOS cameras are arrays of front-side illuminated CCDs. In contrast, the EPIC-pn camera consists of an array of 12 back-side illuminated CCDs, sensitive to energies from 0.15 to 10 keV.  It has the highest soft energy
response of any of the instruments onboard XMM-Newton, and when operated in
timing mode can resolve events down to 0.03 ms.  All of the EPIC-pn data we present
here were obtained in timing mode, necessary because of the high count rate encountered in RS Oph.  This minimized the problem of ``pile-up" in the data (see $\S$3.4 below).  All of the XMM-Newton instruments gather data during each observation, producing simultaneous broadband imaging spectroscopy and high resolution grating spectroscopy data.  The XMM-Newton data were reduced using SAS v. 6.5.  

Before we present and analyze each observation in more detail,
 an overview of the early evolution can be obtained by looking at the
EPIC-pn spectra obtained in timing mode (Fig. 1).  The spectrum became much softer and brighter over the course of the first 40 days.  The total count rate increased by a factor of $\sim$1000 between day 14 and day 54, with most of this increase occuring below 1 keV.  At the same time, the count rate above 2 keV decreased by a factor of 100, demonstrating the dramatic change in hardness ratio of the system.   In light of the X-ray evolution
 of other classical novae (e.g. V382 Vel, Orio et al. 2001a, and 2002) we interpret this behavior
 as follows: the early X-ray emission originated in the ejecta, and as the plasma
 surrounding the white dwarf became optically thinner to supersoft X-rays, the hot
 hydrogen burning white dwarf became visible.  The most striking difference with 
classical novae is the rapidity of this evolution. In the following sections, we show
how the grating spectra support this interpretation, and give unprecedented
 details of the physics of both the ejecta and the inner source. 

\section{Days 13.8 and 26.1: Line Emission from Shock Heated Gas}
RS Oph was observed with X-ray instruments multiple times from day 3 onwards.  Early observations were carried out with the XRT onboard Swift, and the Proportional Counter Array (PCA) on RXTE.  Both detected bright, hard emission from the system.  For the Swift observation on day 3 Bode et al. (2006) obtained a best fit thermal bremsstrahlung temperature of 70 keV (8.1 $\times$ 10$^{8}$ K).  The data obtained with RXTE in the following three weeks showed the spectrum shifting from hard to soft energies, which was interpreted as emission from cooling gas initially shock heated to extremely high temperatures by the blast wave from the nova eruption (Sokoloski et al. 2006).  

\subsection{Day 14}
Both Chandra and XMM-Newton first observed the system 14 days from the start of the outburst, on Feb 26th.   As the source was so hard at this time, the Chandra observation was carried out with the 
High Energy Transmission Grating (HETG) and ACIS-S.  The observation suffered from ``pile-up", a common problem for high count rate observations.  Pile-up is the recording of two or more photons as a single event of higher energy.  Although the use of gratings reduces this problem to some extent, pile-up can still occur for the very brightest objects, and  necessitates some additional reduction steps.  During 
the initial stages of the Chandra reduction pipeline, events with energies greater 
than 15 keV are rejected and flagged as bad.  With grating observations, this leads to 
a ``hole'' in the zeroth order data  i.e. a region with zero counts.  This can result in bad 
centroiding in the source detection phase, creating extraction regions centered on an 
inaccurate zeroth order position.  The consequence is a reduction in the 
accuracy of the wavelength calibration of the spectrum.  To avoid this, we re-extracted 
the spectrum, using new coordinates for the zeroth order image.  We used the 
coordinates of the intersection between the frame-shift streak and the trace of the MEG 
arm in the level 1 event file.  This is the best method available for determining the 
zeroth order position.   This re-extraction ensures the most accurate 
wavelength responses possible are obtained.

Of the suite of XMM-Newton instruments, the EPIC-pn camera and the RGS gratings produced useful data.  The EPIC-MOS instruments were operated in imaging mode in case
there was sudden obscuration of the source like that observed in the classical nova V4743 Sgr (Ness et al., 2003). However, the count rate remained high for the duration of the observation, and the resultant pile-up on the MOS cameras rendered the data unusable. The XMM-Newton and Chandra observations overlap in time substantially, giving combined high spectral resolution from 1.2 to 35 \AA.

The grating data taken on day 14  are shown in Figures 2 and 3.  The total unabsorbed flux in the range covered by each grating, obtained using the ISIS spectral analysis package (Houck \& Denicola, 2000), is given in Table 1. The values range from 1.6 to 5 $\times$ 10$^{-10}$  erg s$^{-1}$ cm$^{-2}$ depending on the energy range covered by each instrument.  In the 5--14 \AA\ region, which is covered by all the grating instruments, the absorbed flux values agree to within 5\%.  Emission lines from H and He-like ions of metals dominate these spectra, and very little continuum emission is present.   We first identified the brightest lines using the GUIDE package, which makes use of the Astrophysical Plasma Emission Database (APED) through ATOMDB v.1.3.  Lines of species from Fe XXV to N VII are present in the wavelength range covered by the combination of the HETG and the RGS.  Where the signal to noise in a line was high enough, we obtained estimates of the centroid velocity and full width at half maximum (FWHM) by fitting a gaussian (for the H-like ions) or sum of gaussians (for the He-like ions) to the data.  The results are presented in Table 2.  For the He-like ions, we constrained the separation of the components and fit the resonance line central wavelength, since this was the strongest line present in each triplet.  An example of this is given in Fig. 4.  Fitting the line profiles yielded a very interesting result - all lines are blueshifted, and the magnitude of the velocity shift increases for lower ionization states and longer wavelengths (see Table 2).  Additionally, as the wavelength increases, so does the broadening of the lines.  

Next, we determined the dominant source of ionization in the plasma using the He-like triplets observed in the spectrum.  We used the ISIS package to convert the count rate spectrum to flux, and then used the best gaussian fit to calculate the flux in each component of the triplet.  Our best fit line fluxes are given in Table 3, along with the value of the so-called G ratio, \textit{(f+i)/r}. This ratio  is a useful plasma diagnostic as its value is sensitive to the dominant source of ionization in the plasma.  Porquet et al. (2001) showed that  \textit{(f+i)/r} is $\sim1$ for collisionally dominated plasmas.  The three lines we obtain ratios for all have \textit{(f+i)/r} = 1 within errors, indicating that at this early stage the plasma is collisionally dominated.

\subsubsection{Modeling the day 14 X-ray emission}

We must bear in mind that the spectra shown in Figures 2 and 3 are unique and unprecedented - no similar spectra have been obtained for any other nova, because in general novae are less X-ray luminous than RS  Oph at this stage of the outburst.  Since the He-like line ratios indicate a collisionally dominated plasma, we attempted to fit the data with the APEC models available in XSPEC, which model line emission from these plasmas.  Following the work done by both Bode et al. (2006)  and Sokoloski et al. (2006) for the prompt X-ray emission, we took the simplest possible approach and first tried to fit a single temperature, solar abundance model to both the Chandra and XMM-Newton spectra simultaneously.  We velocity broadened the model by 1000 kms$^{-1}$ to match the widths observed in the data.  The fits to both datasets are not good, and we find that a single temperature model cannot fit the spectrum.  This is not surprising given that Fe XXV and N VII do not exist at the same temperature---Fe XXV does not exist in plasmas below 10$^{7}$ K, while very little N VII exists above 6.3 $\times$ 10$^{6}$ K (Arnaud \& Rothenflug, 1985).  It is clear that at this stage of the outburst, the X-ray emission is much more complex than a single, thermal brehmsstrahlung model.  Therefore, we decided to add additional temperature components to account for the range of lines seen in the data and improve our model.  This is where one of the advantages of high resolution over broadband data becomes clear; we can make use of the simultaneous detection of both H-like and He-like ions of the same element to determine the range of temperatures required to produce the observed spectrum.  

The ratio of the hydrogen-like Ly $\alpha$ line flux to the sum of the fluxes of the three components of the helium-like triplet is highly sensitive to temperature, and allows us to input real temperature values from the data into our models.  We calculated the ratios for the S, Si, Mg lines in the Chandra spectra, and the O  lines in the RGS spectrum, and present these results in Table 4, along with the temperature derived for each ion.  Using these temperature values, we then fit a multiple temperature component, solar abundance APEC model to the data.  We froze the components with values determined from the line ratios, and included an additional component of higher temperature, (which we allowed to vary) to fit the Fe XXV line observed at 1.8 \AA.  We also allowed the normalization factor, which depends on the X-ray emission measure of the plasma, to vary for each component.  Our final, best fit model gave an absorbing column of 1.2 $\times$ 10$^{22}$ cm$^{-2}$, and four APEC components with temperatures 16.84, 2.31, 0.92 and 0.64 keV.  Introducing additional temperature components did not improve the fit.  The normalization factors suggest X-ray emission measures for each plasma component of a few 10$^{57}$ cm$^{-3}$.  This is approximately a factor of 10 larger than the emission measure determined by Sokoloski et al. (2006) from the RXTE data on the same day in the 0.5--10 keV range.

The fit to the Chandra data is reasonably good, although not perfect.  The continuum is reproduced very well, and the flux in the majority of the lines matches the data to within a factor of 2.  However, the model gives a much poorer fit to the RGS data (see Fig. 3).  Here, the model drastically underestimates the strengths of the O and N lines, by at least a factor of 10 in the case of oxygen, and a factor of 100 for nitrogen.  Introducing additional lower temperature components does not produce the required flux in the O VIII or N VII Ly $\alpha$ lines, and is also inconsistent with the weak O VII and N VI lines observed in the data.  Increasing the normalization factor, which depends on emission measure, produces too much flux at shorter wavelengths.  An explanation for such strong O VIII and N VII lines is enhancement of O and N in the X-ray emitting material.  Livio and Truran (1994) showed that available data suggests that all novae are enhanced in heavy metals, with significant enhancements in CNO cycle nuclei for most ``fast" novae.  Therefore, a likely explanation for the observed line strengths, particularly in the soft part of the spectrum, is that the emission arises in the nova ejecta.  

The model derived fluxes are in good agreement with the ISIS values, even for the RGS data.  Despite the poor fit the O VIII and N VII Ly $\alpha$ lines, these features actually contribute very little to the total X-ray flux from the system at this early stage.  Unabsorbed fluxes were then obtained by correcting for the best fit column density of 1.2 $\times$ 10$^{22}$ cm$^{-2}$.  Assuming a distance of 1.6 kpc, we used these results to calculate a luminosity in the band covered by each instrument.  For the February data, we find a luminosity in the range 5--35 \AA\ of  2 $\times$ 10$^{36}$ erg s$^{-1}$ for the RGS data, 2.78 $\times$ 10$^{35}$ erg s$^{-1}$ in the range 1.5--14 \AA\ for the HEG, and 3.81 $\times$ 10$^{35}$ erg s$^{-1}$ in the range 2.5--25 \AA\ for the MEG.  

In order to have a complete understanding of the observed spectra, we also need to account for the dependence of the velocity shifts and broadening on wavelength and ionization stage. This dependence seems to suggest that the lines are formed in different regions of the nova wind, at different distances from the star.  The same dependence has been clearly observed in another object, the O star $\zeta$  Puppis. Chandra observations of this star showed a similar wavelength dependent behavior of centroid velocity blueshift and broadening to RS Oph (Cassinelli et al., 2001).  There, the authors state that this behavior ``indicates a connection between radial velocity, wind absorption, and Doppler broadening of the line emission".  Recent work by Waldron and Cassinelli (2007) and Cassinelli et al. (2007) showed that a possible origin for X-ray emission in O-star atmospheres is the formation of bow shocks around clumps that form in the winds.  Bow shocks produce a range of post shock temperatures consistent with that seen in O stars, and interestingly, in RS Oph.  Therefore, one possible origin for the line emission seen in RS Oph is the formation of bow shocks in the nova wind, either around clumps in the nova ejecta, or clumps from the red giant wind which are still present in the system (see Crowley, 2006 for a discussion of clumps in the stellar winds of red giants in symbiotic systems).  More complex theoretical models will be required to explore this possibility further.  

\subsection{Day 26}
XMM-Newton observed the system again on day 26 (March 10th). This observation lasted approximately 10000s for the EPIC-pn.  However, the RGS were turned on 2000s after the start of the observation, and were marred by many intervals of high background.  The EPIC-pn count rate in the region 0.15--1 keV has increased by a factor of 2, from 14 to 35 cts/s. while the count rate in the region 1--10 keV has decreased by a factor of 2.  The total absorbed flux in the region 5--35 \AA\ is 1.6  $\times$ 10$^{-10}$  erg s$^{-1}$ cm$^{-2}$, approximately the same as on day 14.  The spectrum is still dominated by emission lines, although the strengths of the lines has changed, reflecting the shift seen in the EPIC-pn data.  The strong lines of Si XIV, Si XIII, Mg XI and Ne X observed in the February spectrum decrease in strength by a factor of $\sim$2, mirrored by a similar drop in total flux in the region 5--14 \AA\ from ~2 $\times$ 10 $^{-10}$  to ~10 $^{-10}$ erg s$^{-1}$ cm$^{-2}$.  In contrast,  O VIII and N VII lines at the soft end of the 
spectrum increase by a factor of $\sim$3, and there is a significant increase in flux between 20 and 30\AA.  The He-like triplet of N VI is also detected, and is at least 10 times brighter than in February, also demonstrating cooling of the emitting plasma.  The temperature indicated by the Mg H/He ratios is 8 $\times$ 10$^{6}$ K, a decrease of 10$^{6}$ K over the course of the 2 weeks since the previous observation.   The total absorbed flux is slightly higher, although very similar to the value obtained in February (see Table 1).

We extracted lightcurves for the observation from both the EPIC-pn and RGS instruments, and found that the count rate was variable, with changes in brightness on times scales of a few hours.   In particular, about 5000s into the observation (3000s for the RGS), a rapid brightening occurred where the count rate increased from $\sim$70 to $\sim $90 cts/s over the course of just a few minutes.  While the flare was detected by both the EPIC-pn and RGS instruments, we show in the upper panel of Figure 5 the EPIC-pn lightcurve extracted in the energy range 0.15 -- 0.4 keV (where the flare was most noticeable) since the EPIC-pn was turned on $\sim$2000 s before the RGS.  This bright period lasted for $\sim$5000s, after which the count rate dropped again, although it remained higher than its initial value for the duration of the observation.  Dramatic changes in X-ray brightness appear to be ubiquitous in novae, and often exhibit detectable periodic oscillations (see e.g. V382 Vel (Orio et al. 2003), V1494 Aql (Drake et al. 2003) and V4743 Sgr (Ness et al., 2005)).  We performed timing analysis on the EPIC-pn and RGS lightcurves.   While we give a detailed discussion in a companion paper (Leibowitz et al., 2007) we summarize the results here as follows.  During the first part of the observation, prior to the flare, no period is found.  Then, for $\sim$5000s after the rapid brightening, a period of 35.6 s is found in the  power spectrum of the observation.  This period then ``turns off" and is not present for the remainder of the observation, even though the count rate remains high.  Similar periods are detected by Swift in the subsequent two weeks (Osborne et al., 2006b).  This XMM-Newton observation is the first detection of this period in the current outburst of RS  Oph.

In order to examine the possible origin of these temporal phenomena in more detail, we split the RGS data into three time periods - before the flare (time interval 1), during the peak of the flare while the $\sim$35 s period is detected (interval 2), and the end of the observation (time interval 3) after the $\sim$35 s period disappears.  We then re-extracted the spectrum for each time interval, applying a less conservative bad interval flagging. We show a comparison of the first and second time periods in the middle and lower panels of Fig. 5.  The spectrum extracted in interval 1(shown in black in the figure) is very similar to February, albeit with different line strengths.  The onset of the flare, however, introduces some dramatic differences into the spectrum.  The second spectrum, is is essentially identical at wavelengths below 20\AA, with the same line shifts, broadening and fluxes.  However, longward of 20\AA\, a dramatic increase in count rate takes place.  The N VI intercombination and forbidden lines of the N VI triplet increase in strength by a factor of 2, although the strength of the resonance line remains the same.  In the region longward of the N VII Ly $\alpha$ line, a large number of new lines with count rates similar to both the N VI and N VII lines suddenly appear.  The lines are still present in the last time interval, once the 35s period disappears, although their count rates are lower reflecting the drop in overall count rate also seen in the lightcurves of the observation.  No significant continuum is detected, and we conclude that these new lines alone are the source of the dramatic increase in count rate at low energies seen in the EPIC-pn data. 

The Ly series lines of C VI lie in this region of the spectrum, and we can identify most of the lines as lines of C VI or N VI if we allow for large blueshifts of 8000--10000 km s$^{-1}$.  However some features, most notably the broad line at 26.7 \AA, remain difficult to identify even allowing for a large velocity shift.  In the lower panel of Fig. 5, we show IDs of the new lines taking the $\sim$8000 km s$^{-1}$ shift into account.  Such velocity shifts are much larger than those observed on day 14, or those of the harder lines detected in the same spectrum.  We note that 8000 km s$^{-1}$ is consistent with the escape velocity of a massive white dwarf.  We find evidence, therefore, that the additional soft flux is due to emission lines appearing in material ejected suddenly at very high velocities.  We also note that the appearance of the new emission lines coincides with the detection of the 35s period in the lightcurve, suggesting that the two phenomena are related.

\section{Days 40, 54 and 67: Hydrogen burning and the supersoft phase}

After day 26, Swift monitoring revealed considerable progress in the trend toward softer energies (Osborne et al. 2006b), so a new observation was obtained with the Chandra LETG/HRC-S on day 40.  The exposure lasted for a little under 10000s (see Table 1)and is shown in black in Figure 6, where we plot the fluxed spectrum for each of the SSS phase observations.  Since the fluxed spectra are corrected for the instrumental response, they allow a comparison between datasets obtained with different instruments.  The day 40 spectrum is markedly different from that obtained just two weeks earlier with XMM-Newton.  The total absorbed flux in the 5--83 \AA\ region is now 1.7 $\times$ 10$^{-9}$ erg s$^{-1}$ cm$^{-2}$, a factor of $\sim$100 larger than the day 26 observation when calculated in the same wavelength range.  Longward of 14 \AA, a dramatic increase in the level of the continuum has occurred.  Several absorption features are present which are predicted by white dwarf model atmospheres, most notably lines of Fe XVII, O VIII and N VII.  The spectrum is similar to that observed in V4743 Sgr during its supersoft phase, although the RS Oph spectrum is significantly harder.  Also, as far as we can determine, there is no significant blueshift of the absorption lines as was the case for V4743 Sgr.  We also detect lines due to O I at 23.51 \AA\ and O II at 23.35 \AA\ that are known to be associated with absorption by cold gas ( e.g. Yao \& Wang, 2006), which we attribute to absorption by interstellar or circumstellar material.

In addition to the absorption features, the emission line features observed on days 14 and 26 are still present across the spectrum in Figure 6.  In the harder portion of the spectrum (between 5 and 14 \AA), we still detect several of the emission lines that were prominent in the previous spectra, most notably the lines of Mg XII, XI and Ne X at 8.4, 9.1 and 12.1 \AA\ respectively.   The lines have continued to fade, and the total flux in the 5--14 \AA\ is now 6.8 $\times$ 10$^{-11}$ erg s$^{-1}$ cm$^{-2}$ --- 30\% lower than in the XMM-Newton observation of day 26.1.  In the softer part of the spectrum, emission lines of Fe XVII, O VIII, O VII, N VII and N VI that arise in the nebula are present, superimposed on the continuum.

We extracted the lightcurve for the observation using the CIAO tool \textit{lightcurve} (see Fig. 7, also Leibowitz et al., 2007), and found that RS  Oph was still highly variable.  Unlike the XMM-Newton observation 2 weeks earlier, the Chandra data show large scale fluctuations in brightness throughout the entire observation, with no obvious extended ``bright'' period.  We did extract a separate spectrum for the period marked by the dotted vertical lines in Fig 8, since during this period the lightcurve is relatively flat.  However, we found no evidence of a dramatic softening of the spectrum as we did for the EPIC-pn data earlier in the month.  The flat region, compared to the rest of the spectrum, shows a higher count rate but no evidence of a dramatic shift in energy distribution.  The change in count rate is confined to the wavelength range 14--30 \AA.  However, since the shape of the spectrum and the absorption features do not seem to change, the variability may be due to changing absorption in the clumpy ejecta.

Two more observations were carried out in the subsequent 4 weeks.  The first was with XMM-Newton on day 54.  Again, only the RGS and EPIC-pn obtained useful information, and even then both instruments suffered from pile-up problems due to very 
high count rate.  This effect has been encountered in another observation of a bright nova, V4743 Sgr in 2003 (Leibowitz et al., 2006).  In the RGS, pile-up results in 
first order events being incorrectly identified as higher order.  This results in a loss of 
signal in the first order spectrum.  This can be corrected to some extent by examining 
the pulse height amplitudes (PHA) of the higher order events.  Piled up events 
increase in PHA by an integer multiple of the intrinsic energy.  Therefore, events with 
unreasonably large PHA in higher orders can be identified as lower order pile-up 
artifacts.  A correction can then be applied, where the higher order counts are 
multiplied by the order and added to the first.  This gives the corrected first order count 
distribution.  This works very well for extremely soft sources, as no intrinsic higher 
order counts are present.  Therefore, all higher order events can be confidently 
identified as artifacts of pile-up.  However, this correction proved to be more difficult for 
RS Oph than for V4743 Sgr, as the source is harder and therefore has real higher order events.  The 
softest end of the second order event distribution is intrinsic to the source, while the 
hard second order events are due to pile-up.  Therefore, the correction to first order is 
only valid within a certain wavelength range, which has to be estimated.  We therefore 
lose the softest part of the first order RGS spectrum, although we have regained the 
counts at the hard end, and we end up extracting the spectrum in the region 5--28 \AA.  Although this correction sacrifices some spectral coverage, 
none of the RGS orders would be suitable for model fitting if this procedure had not 
been carried out.

As Fig. 1 demonstrated, by the XMM-Newton observation on day 54, RS Oph had become an incredibly bright, supersoft source. The RGS spectrum (shown in green in Fig. 6) is similar in shape to the LETG spectrum of two weeks earlier, although it has become even brighter, reflecting the huge increase in count rate seen in the broadband data.  The total absorbed flux can only be evaluated in the range 5--28 \AA, and because of the method used to correct for pile-up, we could not use ISIS to estimate the flux.  Instead, we rely on the fit to the model described below, and estimate the flux in the region 5--28 \AA\ to be at least 4 $\times$ 10$^{-9}$ erg s$^{-1}$ cm$^{-2}$ .   This is a factor of 2.5 higher than the LETG observation of day 40 when evaluated in the same wavelength range. The absorption features are deeper, and the emission features appear to be less prominent.  The Ly series of N VII in absorption is now easily identifiable between 19 and 25 \AA.  The neutral oxygen absorption features are also more pronounced.  The harder lines between 5 and 14 \AA\ are still present, and once again are fainter than the previous observation.   We extracted the lightcurve for  the EPIC-pn and both RGS instruments, and found that the large scale brightness fluctuations observed two weeks earlier with Chandra are no longer present.  Instead, for the duration of the observation the count rate remains constant.

Another observation was carried out using Chandra on day 67, again with the LETG/HRC-S, with a net exposure time of $\sim$6500s.  The spectrum (shown in red in Fig. 6) is remarkably similar to that observed with XMM-Newton on day 54.  The total absorbed flux measured with ISIS is 3.8 $\times$ 10$^{-9}$ erg s$^{-1}$ cm$^{-2}$ in the range  5--83  \AA.  In the same range as the corrected RGS spectrum (5--28 \AA) we find that the absorbed flux is 3.4 $\times$ 10$^{-9}$ erg s$^{-1}$ cm$^{-2}$ --- about 15\% lower than the day 54 observation.  Again, it appears that the absorption features are less contaminated by superimposed emission, although both the harder lines between 5 and 14 \AA\, and the N VI He-like triplet are clearly detected, showing that some nebular emission is still present at this time.  The hard nebular component has faded further, and the total 5--14 \AA\ flux is now just 3.5 $\times$ 10$^{-11}$ erg s$^{-1}$ cm$^{-2}$.  Extracting the lightcurve, we again found no large scale fluctuations in brightness, with the lightcurve looking similar to one obtained with XMM-Newton two weeks earlier.

The $\sim$35 s period is missing on day 40, but a period of 17.86 s is clearly detected in the data.  In a companion paper by Leibowitz et al. we present a discussion of the possible origin of this period, that seems to be due emission originating near the poles of the white dwarf which are at different temperatures.  If this interpretation is correct, the poles must have reached a uniform temperature by day 54, as a period of 34.72s was detected in the EPIC-pn and RGS data.  Later, no clear period is identified in the LETG data obtained on day 67.  Since on days 54.0 and 67 no large scale fluctuations were detected in the observation lightcurves,  the appearance of the $\sim$35 period is at this point no longer associated with increases or decreases in the lightcurve or with spectral variability. 

\subsection{Determining the white dwarf temperature: NLTE atmospheric models}

The dramatic increase in the continuum level between 14 and 35 \AA\ in these spectra allowed us to attempt to fit white dwarf atmospheric models to the data and investigate the temperature of the star.  Many of the Non Local Thermonuclear Equilibrium (NLTE) models commonly used for extremely 
hot white dwarfs and central stars of planetary nebulae, for example, those developed by Hartmann \& Heise, (1997) are not suitable because they do not reach the extremely high temperatures of post-nova white dwarfs. We used a series of plane parallel, static models calculated by Rauch (2003)\footnote{Most of the Rauch models are public, and can be downloaded from http://astro.uni-tuebingen.de/~rauch/}.  These models extend up to 10$^{6}$ K and include an extensive network of elements from H to Fe for line blanketing calculations to produce state of the art atmospheric spectra.  We use models with two sets of abundances, solar or halo (which were developed for modeling systems in the Magellanic Clouds).  

We also used a series of models developed specifically for the classical nova V4743 Sgr (Rauch et al., 2005a,b), with abundnaces which follow the composition of the white dwarf through multiple cycles of accretion and nova outbursts (Rauch, private communication).  Nuclear processing and dredge-up of the underlying white dwarf lead to surfaces that are enhanced in He and CNO cycle elements.  Nuclear processing of the accreted material transforms C into N via the CNO bi-cycle, leading to a C/N ratio that is significantly lower than solar.  This type of model fitting allows us to explore the surface composition and determine whether the white dwarf shows signs of having retained its CNO processed accreted material.  If this is the case, the white dwarf can grow in mass and possibly explode as a type 1a supernova.  A coarse grid of abundances is available, allowing the C/N ratio to be varied from the solar value, to 0.001 times solar.   We note that these abundances have not been fine-tuned for the RS  Oph system, but at the present time they are the only suitable models available to use in the literature.  By showing that even with these models useful constraints are obtained, we hope to encourage further development of white dwarf atmospheric models specifically for use with RS  Oph.

We constrained our fits to the region longward of 14 \AA\ where the white dwarf continuum is strongest, and to avoid the harder lines of the nebular emission.  For the Chandra observations we modeled both the +1 and -1 orders simultaneously, and for the XMM-Newton observation both RGS1 and 2.  First, we fit all available models to each observation in order to explore the dependence of best fit photospheric temperature on abundance.  Since the atmospheric abundances are uncertain, and our nova models are not finely tuned to the current dataset, this was necessary to help estimate the uncertainty in our best fit temperature, and absorbing column.  With the exception of the halo model, all the temperatures found agree to within $\sim$10,000 K.  The halo model was developed for novae in the LMC and SMC, and since optical and infrared studies of the RS Oph system seem to indicate that the red giant has near solar abundances, we believe this is an inappropriate model to consider for RS Oph. Therefore, we believe the temperature that model fitting returns is relatively independent of the choice of abundances.  

Next, we determined the best fit model.  We did this primarily by eye, since the presence of emission features arising in the nova ejecta, and absorption lines due to the ISM/CSM, make determining a formal statistical fit difficult, especially without a fine tuned model.  The fit to the day 40 data obtained with Chandra is the worst of the three, due to the superposition of many emission features on the white dwarf continuum, and we cannot determine a reliable temperature for this observation (the fit is however useful for obtaining lower limits to the flux, as we show below).  However, as the continuum continues to increase in brightness, and the superimposed emission features become weaker, the spectral fits improve.  On day 54, the best fit is obtained with the V4743 Sgr model which has a C/N ratio of 0.001 times the solar value.  This model is necessary to explain the lack of a C VI K-edge at 25.37 \AA\ and weak C VI Ly series lines between 26 and 30 \AA\ in the observed spectrum.  Our best fit gives a temperature of $\sim$820,000 $\pm$ 10,000 K, and an absorbing column of 2.3 $\times$ 10$^{21}$ cm$^{-2}$.  While this value is slightly lower than the interstellar value of 2.4$\times$10$^{21}$ cm$^{-2}$ (Hjellming et al. 1986), we believe it to be a lower limit, since the lack of data longward of 28 \AA\ inhibits our ability to fit this parameter with a large degree of confidence.

The Chandra data of day 67 provides the most convincing evidence of a high temperature for the white dwarf photosphere.  The continuum level is so high that the contribution from the emission component is almost insignificant over most of the spectrum.  The observation is also free of the instrumental issues that affected the XMM-Newton observation two weeks earlier.  We include two figures showing the fit to the day 67 data.  The first, Figure 8, shows our best fit model with a blackbody for comparison.  This figure clearly demonstrates the importance of NLTE effects in modeling the spectra of these hot white dwarfs.   Figure 9 demonstrates the difference between our nova model with C/N = 0.001 solar and a model of solar abundance.  Again, no C VI K-edge or Ly series lines are detected in the tail of the continuum, showing clearly that our nova abundance model is more appropriate than the solar abundance model.  We note that the O K-shell edge at 23.4 \AA\ is partially filled, which seems to be due to superimposed emission.  For day 67, we find a best fit temperature of 809,000 $\pm$ 10,000 K and an intervening absorbing column of 3.16 $\times$ 10$^{21}$ cm$^{-2}$.  This is slightly lower than the value obtained for day 54, showing that the white dwarf has started to cool.  The day 67 best fit value of N(H) is higher than the interstellar value.  Although a large uncertainty is associated with these values because of the poor fit at shorter wavelengths, constraining N(H) to the interstellar value produces model fits with an excess of emission at long wavelengths.  Therefore we conclude that there is still some intrinsic absorbing material present in the system.

Using the best fit model to the day 40 data, and correcting for the intervening absorbing column, we derive a value for the luminosity in this range, L$_{5-83 {\rm \AA}}$ = 8.1 $\times$ 10$^{36}$ erg s$^{-1}$.  Given the difference between the ISIS derived flux (which is essentially the directly measured flux and includes the contribution from the nebula) and the flux from the model (which is only for the white dwarf atmosphere), we consider this luminosity a lower limit - the true luminosity could be as much as 20\% higher, since here we assume all additional flux arises in the nebula.  On day 54, we derive a value for L$_{5-28 \rm {\AA}}$ =  5.8 $\times$ 10$^{36}$ erg s$^{-1}$.  Here, we can also use the fit to the EPIC-pn data to give us additional information.  Using the same model, we find a lower limit to the absorbed flux in the energy range 0.15--10 keV of $\sim$3 $\times$ 10$^{-9}$ erg s$^{-1}$ cm$^{-2}$, and L$_{0.15-10 \rm {keV}}$ = 9.2 $\times$ 10$^{37}$ erg s${-1}$. The luminosity value determined on day 67 is a little lower;  L$_{5-83 \rm{ \AA}}$ = 3.31 $\times$ 10$^{37}$ erg s$^{-1}$.  We estimate the errors on our luminosity to be between 10 and 20\% because of the nebular contribution which the atmosphere model does not account for.  

All the models we fit to the data had log(g) = 9, as these were the only models that could produce a high enough temperature to match the data.  Although the grid of effective gravity value of Rauch's models is coarse, models with g=7, 8 and 9 are available.  Obviously the error in the log(g) determination is large, but fitting models with different abundances that the effective temperature is determined in a relatively narrow range and even fitting a blackbody we obtain a temperature close to 800,000 K.  We can express the surface gravity in terms of temperature assuming that the white dwarf luminosity does not exceed (at least not by much and not for a long time) the Eddington luminosity given by 
\begin{equation}
L_{\rm {Edd}} \simeq 1.3 \times 10^{38} \hspace{2mm} \frac{M_{WD}}{M_\odot} \hspace{2mm} \rm {erg \hspace{1mm}s^{-1}}
\end{equation}Then, substituting r$^{2}$ = GM/g into the expression L = 4$\pi$r$^{2}$$\sigma$T$^{4}$, and equating to the expression for the Eddington luminosity above, we can write
 \begin{equation}
 g \geq  3  \times 10^{8} \hspace{2mm} \left( \frac{ T (K)}{800000} \right)^4 \hspace{2mm} { \rm cm \hspace{1mm}s^{-2}}
\end{equation} Therefore, the main constrain on the effective gravity is determined by the high effective temperature, and it appears that in our case g must be at least few times 10$^8$ cm s$^{-2}$. A less compact WD would result in a bolometric luminosity exceeding the Eddington limit by a large amount.

\section{June, September and October: The end of surface burning and the road to 
quiescence}
By early June, RS Oph was no longer detected as a luminous supersoft source.  Swift observations obtained during April and May showed a linear decline in count rate in the 0.3--10 keV region, from a 
maximum of $\sim$400 c/s on April 12, to 0.3 c/s by May 31st.  Between May 31 and 
June 11, there were no significant changes in this count rate (Osborne et al, 2006c,d).  Chandra observed the system again on June 4th (day 111), 6 weeks after the 
previous grating observation.    The data confirm the huge drop in flux seen in the Swift observations.  While the day 67 spectrum had a high count rate and strong continuum emission, the day 111 spectrum (shown in the upper panel of Figure 10) is dominated by emission lines with only a low continuum level.  The count rate has also decreased by a factor of $\sim$170, to only 0.22 c/s.  We obtained the absorbed flux in the region 5--35 \AA\ using ISIS and obtained a value of 2 x 10$^{-11}$ erg s$^{-1}$ cm$^{-2}$.  It seems clear from the dramatic change in the system from the end of April that surface hydrogen burning has stopped.  

Where the signal to noise was high enough, we estimated spectral line velocity shifts by fitting a gaussian model.  The Mg XII line at 8.42 \AA\, the Fe XVII line at 15.01 \AA\ and the Ly $\alpha$ lines all show evidence of a blue shift of $\sim$ 500 kms$^{-1}$.  None of the other lines showed a detectable shift, and within the uncertainties of the model fit all the observed lines could be at their rest wavelengths.  This is in stark contrast with the huge blueshifts observed in February and March.  The spectra obtained in September (day 205) and October (day 239) with the RGS are very similar to the June LETG data.  Figure 11 shows the two spectra.  Again, the dramatic blue shifts of the emission lines observed in March and April are no longer present, and instead lines are detected at their rest wavelengths. There is little change between the two, which were obtained 4 weeks apart.  The count rate measured by EPIC-pn in the region 0.15--10 keV on these dates falls from 0.76 c/s to 0.59 c/s.  We attempted to fit a variety of models to these data sets, and once again did not achieve a good fit.  The absorbed fluxes for both observations are given in Table 1, and show that the luminosity has decreased by a factor of $\sim$10 since June.  Despite the lack of spectral resolution, it is still clear from the EPIC-pn data that much of the emission is from lines.  Interestingly, emission from Fe XXV at 6.7 keV, and the Fe K$\alpha$ complex  at $\sim$6.4 keV is still detected in September (see Fig. 12), although the count rate is a factor of $\sim$1000 lower than in February, when the Fe XXV He-like triplet was detected with Chandra and the HETG, and also with EPIC-pn on XMM-Newton.  By October, the count rate in this region has become too low for a confident identification of the Fe features.

We attempted to fit the data with various thermal plasma models available in XSPEC, and did not achieve a good fit with any of them.   We also tried to fit the data with models suitable for accreting CVs, including a cooling flow and models developed for photoionized plasmas, and once again could not obtain a good fit.  The photoionized plasma models, typical for most magnetic CVs (Mukai et al., 2003), predict a stronger continuum especially at hard energies, than we see in the data.   The cooling flow models  for accretion disks (Mukai et al., 2003)  fail to reproduce the line ratios observed in the data, and we note that the O VIII Ly $\alpha$, N VII Ly $\alpha$ and Fe XVII lines between 15 and 24 \AA\ were particularly hard to reproduce. 

For the observations in September and October, conversion of the EPIC-pn count rate to an equivalent ROSAT PSPC count rate shows that the X-ray flux is still at least an order of magnitude higher than in quiescence (Orio 1993, Orio et al. 2001b).  The only similar spectrum observed in a nova that of the recent nova, V382 Velorum (Nova Velorum 1999).  V382 Vel provides a very interesting comparison system to RS  Oph.  Ness et al., (2005), present a Chandra observation of the system also obtained around 8 months after the outburst.  The authors observed a similar emission line dominated spectrum which they attributed to emission in a collisionally ionized and excited shell.  The V382 Vel LETG spectrum is included for comparison in the lower panel of Fig. 10. The spectrum of RS Oph has stronger 
lines in the region 10--30 \AA, particularly those of O VIII Ly $\alpha$ and N VII Ly $
\alpha$.  The Fe XVII lines between 15 and 17 \AA\ are present and bright, unlike in 
V382 Vel where their absence was suggested to imply Fe under-abundance by Ness 
et al. The lines at longer wavelengths seen in V382 Vel are not present in RS Oph in 
June despite a similar absorbing column, suggesting the plasma is hotter at this stage.  However, these lines emerge in the RGS spectra obtained in  September and October.

A similar spectrum was also seen during the ``low" state of the V4743 Sgr LETG observations discussed in \S3.  In that observation, the white dwarf appears to have been suddenly obscured, and in the place of the strong supersosft continuum, only a very soft emission line spectrum was observed showing lines of N VII and N VI (Ness et al, 2003).  The similarity of these spectra make it 
likely that they share a common emission mechanism, most likely originating in the 
material ejected by the novae.  

\section{Discussion}
The appearance of additional plasma at low energies on day 27, associated with very high velocity material, is extremely interesting.  Its appearance may be related to the jet-like structure detected in radio data (see O'Brien et al., 2006, Lane et al., 2007).  If these new lines are associated with the jet-like outflow, then the absence of a system of red-shifted lines means we are only seeing one side of the jet at this time.  

The high effective temperature of the WD determined at the end of March and in April implies an effective gravity of at least a few times 10$^{8}$ cm s$^{-2}$.  As discussed previously, only models with log(g) = 9 can produce any flux shortward of $\sim$16 \AA.  Therefore, in order for the observed luminosity not to be extremely super-Eddington, the white dwarf must be extremely compact, and therefore massive.  Early work by Paczy\'nski (1971) on the evolution of planetary nebulae nuclei showed that only the most massive cores, those above 1.2 M$_\odot$, ever reached temperatures of log(T$_e$) $>$ 5.7 during their shell burning phase.  Sequences of white dwarf configurations for hot, massive white dwarfs were calculated by Althaus et al. (2005) for various effective temperatures and chemical compositions.  Although their grids only reach 150,000 K, their results clearly indicate that only the most massive white dwarfs with M$_{WD}$ $\geq$ 1.18 M$_{\odot}$ can reach log(g) $\geq$ 9 for any temperature.  Therefore, both our high photospheric temperature, and large surface gravity, only appear to be consistent with a massive white dwarf, of at least 1.2 M$_\odot$, and values larger than this seem likely.    

Swift monitoring of RS Oph in April and May (Osborne et al., 2006d) showed that after April 28.7 (day 75 of the outburst) the soft X-ray flux began to fade, with an almost linear decline in count rate throughout May.  Our data obtained on day 67 with the LETG (which is more sensitive to the soft part of the spectrum than Swift) show that the white dwarf was already less luminous and most likely cooler on day 67 than on day 54.  We attribute this drop in luminosity to the end of nuclear burning.  Hachisu and Kato (2007) assumed that hydrogen burning ended on day 75, and from model fitting determined that the mass of the white dwarf was 1.35 M$_{\odot}$.  However, given that burning duration is inversely proportional to white dwarf mass, we conclude from our data that the white dwarf mass may be even higher than this estimate.  This places RS Oph extremely close to the Chandrasekhar limit.  Hachisu and Kato also predicted a continuum driven wind phase lasting 42 days for a model with hydrogen fraction X = 0.17.  We note that the lightcurve of RS Oph was extremely variable in March, both in the XMM-Newton and Chandra observations.  We examined archival observations obtained with Swift during this period, and although many of the exposures are only a few thousands seconds in length, we also find large variability in those data.  However, in April the lightcurve becomes almost stable, with no large changes in brightness.  If all variability phenomena observed in March are associated with the evolution of the nova ejecta, as we see in our observation on day 27, we can conclude that mass loss turns off at the end of March, which is consistent with Hachisu and Kato's' prediction that mass loss ends on day 42  - firmly between our day 39 and day 54 observations.

\section{Conclusions}

RS Oph was detected in outburst on February 12th, 2006.  The evolution of the outburst during the first 9 months, as seen with high resolution grating spectra, can be summarized as follows.
The spectra are initially dominated by H-like and He-like lines of Fe XXV through N VII, produced as 
the nova ejecta interact with the red giant wind filling the RS  Oph system.  The lines are blue-shifted by velocities between 500 and 1300 km s$^{-1}$, and broadened by $\sim$2000 km s$^{-2}$.  The strengths of the O and N lines are consistent with the ejecta being enhanced in these elements, as seen in many other novae.  On day 27, additional plasma emitting lines in the region 24--30 \AA\ appears $\sim$5000 s into the observation.  Many of these features are consistent with lines of C VI and N VI blue-shifted by $\sim$8000 km s$^{-1}$.  If this is the case, they could be associated with the outflow observed in the VLBI radio data obtained just a few days later.  Temperatures obtained from H-like to He-like line ratios indicate multiple temperatures between 10$^6$ and 10$^7$ K, with substantial cooling between the two observations.  

The system then evolves rapidly toward a softer state, as the nebular component continues to cool 
and the underlying white dwarf atmosphere begins to emerge. At the end of March and in April, RS Oph appears as a bright supersoft source.  We fitted NLTE atmospheric 
models in order to determine some physical parameters of the white dwarf.  We find 
log(g) $\sim$9, and a photospheric temperature of $\sim$800,000 K.   The best fits are obtained with models which have C/N abundance ratios of at most  0.1 times the solar value, indicative of CNO cycle processed material on the WD surface, which may indicate that some mass is retained after the accretion-outburst cycle.  The spectral changes observed from days 39 to 69 are consistent with mass loss ending at the end of March, and nuclear burning turning off at the end of April.

Just 8 weeks later, the bright 
supersoft emission from April has been replaced in June by an emission line 
dominated spectrum with very little continuum.  Many of the same lines observed in February and March are still detected, although the large velocity shifts have disappeared.  The spectrum is similar to that 
seen in other post novae, including V382 Vel and V4743 Sgr, although the range of 
lines and inferred temperatures is different.  In those systems, the emission is believed 
to arise in the nova ejecta, and is most likely the case for RS Oph.  We compared the late spectra to those of several accreting CVs, and conclude that an accretion disk is not the source of the X-ray emission.  The system has still not returned to its quiescence level in X-rays, exhibiting a flux at least 10 times higher than the maximum quiescence flux measured with ROSAT.  Continued monitoring in both the X-ray and optical regimes will be crucial in understanding the behavior of this system as it begins to accrete again and return to its quiescent state.

We are extremely grateful to Harvey Tananbaum, the director of Chandra, and to Norbert Schartel, the ESA XMM-Newton Project Scientist, for their generous allocation of discretionary observing time.  We thank the anonymous referee for their useful and insightful comments.  This research was supported by
NASA grant for XMM-Newton research, by Chandra-Smithsonian award, and by an Italian Space Agency (ASI) grant 2006 for X-ray astronomy data analysis.  The latter provided partial salary support to TN.  This project was completed while TN and MO visited the Kavli Institute for Theoretical Physics (KITP) in Santa Barbara, and MO gratefully acknowledges the financial support of KITP.

\clearpage
\thispagestyle{empty}
\begin{deluxetable}{ccccccc}
\tabletypesize{\scriptsize}
\rotate
\tablecaption{Summary of Observations\label{tbl-1}}
\tablewidth{0pt}
\tablehead{
\colhead{Date\tablenotemark{a}} & \colhead{Facility} & \colhead{Instrument} & 
\colhead{Exposure time (s)} & \colhead{Days from Outburst\tablenotemark{b}} &\colhead
{Count Rate (cts/s)\tablenotemark{c}} & \colhead{Absorbed Flux (erg s$^{-1}$ cm$^{-2}$})
}
\startdata
02/26/2006 15:18:49 & Chandra & HETG/ACIS-S MEG & 9912 & 13.81 & 3.34 $\pm$ 2.75 
$\times$ 10$^{-2}$ & 3.3 $\times$ 10$^{-10}$ (2.5--25 \AA) \\
& & HEG & 9912 & & 1.63 $\pm$ 2.36 $\times$ 10$^{-2}$ & 4.1 $\times$ 10$^{-10}$ (1.5--14 \AA) \\
02/26/2006 16:51:47 & XMM-Newton & EPIC-pn & 21411 & 13.87 & 91.64 $\pm$ 
6.77 $\times$ 10$^{-2}$ & 5.0 $\times$ 10$^{-10}$ (0.15--10 keV) \\
& & RGS1 & 23838 & & 1.85 $\pm$ 9.35 $\times$ 10 $^{-3}$ & 1.6 $\times$ 10$^{-10}$ (5--35 \AA)\\
& & RGS2 & 23833 & & 2.81 $\pm$ 1.14 $\times$ 10 $^{-2}$ & 1.6 $\times$ 10$^{-10}$ (5--35 \AA)\\
03/10/2006 23:03:39 & XMM-Newton & EPIC-pn & 10358 & 26.13 & 73.01 $\pm
$ 8.86 $\times$ 10 $^{-2}$ & 2.3 $\times$ 10$^{-10}$ (0.15--10 keV)\\
& & RGS1 & 1142 & & 1.36 $\pm$ 3.54 $\times$ 10 $^{-3}$ & 1.6 $\times$ 10$^{-10}$ (5--35 \AA)\\
& & RGS2 & 1119 & & 2.63 $\pm$ 4.95 $\times$ 10 $^{-2}$ & 1.6 $\times$ 10$^{-10}$ (5--35 \AA)\\
03/24/2006 12:24:18 & Chandra & LETG/HRC-S & 9969 & 39.69 & 17.24 $\pm$ 5.41 
$\times$ 10$^{-2}$ & 1.7 $\times$ 10$^{-9}$ (5--83 \AA)\\
04/07/2006 20:47:51 & XMM-Newton & EPIC-pn & 3850 & 54.04 & 1947 $\pm$ 
0.71 & 3.0 $\times$ 10$^{-9}$ (0.15--10 keV)\\
& & RGS1 & 9802 & & 165 $\pm$ 0.14 & 4.0 $\times$ 10$^{-9}$ (5--28 \AA)\\
& & RGS2 & 18640 & & 131 $\pm$ 9.33 $\times$ 10 $^{-2}$ & 4.0 $\times$ 10$^{-9}$ (5--28 \AA)\\
04/20/2006 17:23:48 & Chandra & LETG/HRC-S & 6523 & 66.89 & 38.95 $\pm$ 9.43 $
\times$ 10$^{-2}$ & 3.8 $\times$ 10$^{-9}$ (5--83 \AA)\\
06/04/2006 12:04:54 & Chandra & LETG/HRC-S & 19966 & 111.67 & 0.22 $\pm$ 1.57 
$\times$ 10$^{-2}$ & 2.0 $\times$ 10$^{-11}$ (5--83 \AA)\\
09/06/2006 01:41:10 & XMM-Newton & EPIC-pn & 25552 & 205.24 & 0.756 $
\pm$ 5.94 $\times$ 10 $^{-3}$ & 1.2 $\times$ 10$^{-12}$ (0.15--10 keV) \\
& & RGS1 & 40257 & & 3.139 $\times$ 10$^{-2}$ $\pm$ 1.52 $\times$ 10 $^{-3}$ & 9.1 $\times$ 10$^{-13}$ (5--35 \AA)\\
& & RGS2 & 40247 & & 4.306 $\times$ 10$^{-2}$ $\pm$ 1.62 $\times$ 10 $^{-3}$ & 9.1 $\times$ 10$^{-13}$ (5--35 \AA)\\
10/09/2006 23:20:16 & XMM-Newton & EPIC-pn & 43339 & 239.14 & 0.585 $
\pm$ 4.00 $\times$ 10 $^{-3}$ & 9.1 $\times$ 10$^{-13}$ (0.15--10 keV) \\
& & RGS1 & 48753 & & 1.938 $\times$ 10$^{-2}$ $\pm$ 1.28 $\times$ 10 $^{-3}$ & 5.38 $\times$ 10 $^{-13}$ (5--35 \AA)\\
& & RGS2 & 48747 & & 3.045 $\times$ 10$^{-2}$ $\pm$ 1.39 $\times$ 10 $^{-3}$ & 5.38 $\times$ 10 $^{-13}$ (5--35 \AA)\\
\enddata
\tablenotetext{a}{Date format MM/DD/YYYY hh:mm:ss}
\tablenotetext{b}{t$_{0}$ = 2453779.33 after Narumi (2006)}
\tablenotetext{c}{Chandra count rates are for first order only}
\end{deluxetable}
\clearpage

\begin{deluxetable}{ccccc}
\tablecaption{Centroid velocities and FWHM of February emission lines}
\tablewidth{0pt}
\tablehead{
\colhead{Ion} & \colhead{Grating} &\colhead{log Peak emissivity (K)} & \colhead{Centroid velocity (km s$^{-1}$)}  & \colhead{FWHM (km s$^{-1}$)}
}
\startdata

 S XV & MEG & 7.2 & -595 $\pm$ 178 & 1741 $\pm$ 119\\
 Si XIV & MEG & 7.2 & -496 $\pm$ 56 & 1776 $\pm$ 49\\
 Si XIII & MEG & 7.0 & -632 $\pm$ 45 & 1403 $\pm$ 45\\
 Mg XII & MEG & 7.0 & -926 $\pm$ 73 & 2022 $\pm$ 36\\
 Mg XI & MEG & 6.8 & -1145 $\pm$ 74 & 1918 $\pm$ 99\\
 O VIII & RGS2 & 6.5 & -1202 $\pm$ 47 & 2029 $\pm$ 48\\
 N VII & RGS2 & 6.3 & -1307 $\pm$ 72 & 2108 $\pm$ 73\\
\enddata

\end{deluxetable}
\clearpage

\begin{deluxetable}{ccccc}
\tablecaption{He-like triplet line fluxes and ratios}
\tablewidth{0pt}
\tablehead{
\colhead{Ion} & \colhead{F$_{\textit{r}}$\tablenotemark{a}} &\colhead{F$_{\textit{i}}$\tablenotemark{a}} & \colhead{F$_{\textit{f}}$\tablenotemark{a}}  & \colhead{\textit{(f+i)/r}}
}
\startdata
S XV & 2.01$\pm$ 0.27 &  0.43 $\pm$ 0.24 & 1.18 $\pm$ 0.18 & 0.8 $\pm$ 0.23\\
Si XIII & 1.79 $\pm$ 0.09 & 0.58  $\pm$ 0.07 & 1.35 $\pm$ 0.07 & 1.08 $\pm$ 0.07\\
Mg XI & 1.22$\pm$ 0.21 &  0.37 $\pm$ 0.19 & 0.72 $\pm$ 0.12 & 0.9 $\pm$ 0.26\\
\enddata

\tablenotetext{a}{Flux in units of 10$^{-14}$ erg s$^{-1}$ cm$^{-2}$}
\end{deluxetable}
\clearpage

\begin{deluxetable}{rrrrrr}
\tablecolumns{6}
\tablewidth{0pc}
\tablecaption{H/He-like ratios on days 14 and 27}
\tablehead{
\colhead{} & \multicolumn{2}{c}{Day 12.3} & \colhead{} & \multicolumn{2}{c}{Day 27.7}\\
\cline{2-3} \cline{5-6} \\
\colhead{ion} & \colhead{ratio} & \colhead{T (K)} & \colhead{} & \colhead{ratio} & \colhead{T (K) }
}
\startdata
& & & & & \\
\sidehead{Chandra\tablenotemark{a}}
S & 0.48$\pm$0.17 & 1.6 $\times$ 10$^{7}$ & & \nodata & \nodata \\
Si & 3.16$\pm$0.04 & 2.0 $\times$ 10$^{7}$ & & \nodata & \nodata \\
Mg & 1.06$\pm$0.11 & 7.9 $\times$ 10$^{6}$ & & \nodata & \nodata \\
\sidehead{XMM-Newton}
Mg & 1.31$\pm$0.08 & 8.9 $\times$ 10$^{6}$ & & 0.65$\pm$0.10 & 6.3 $\times$ 10$^{6}$ \\
O & 8.70$\pm$0.14 & 5.6 $\times$ 10$^{6}$ & & \nodata & \nodata \\
\enddata

\tablenotetext{a}{All values quoted are taken from the MEG data, which had significantly higher signal to noise in each line than the HEG.}
\end{deluxetable}

\clearpage

\begin{figure}
\begin{center}
\includegraphics[width=5in,angle=270]{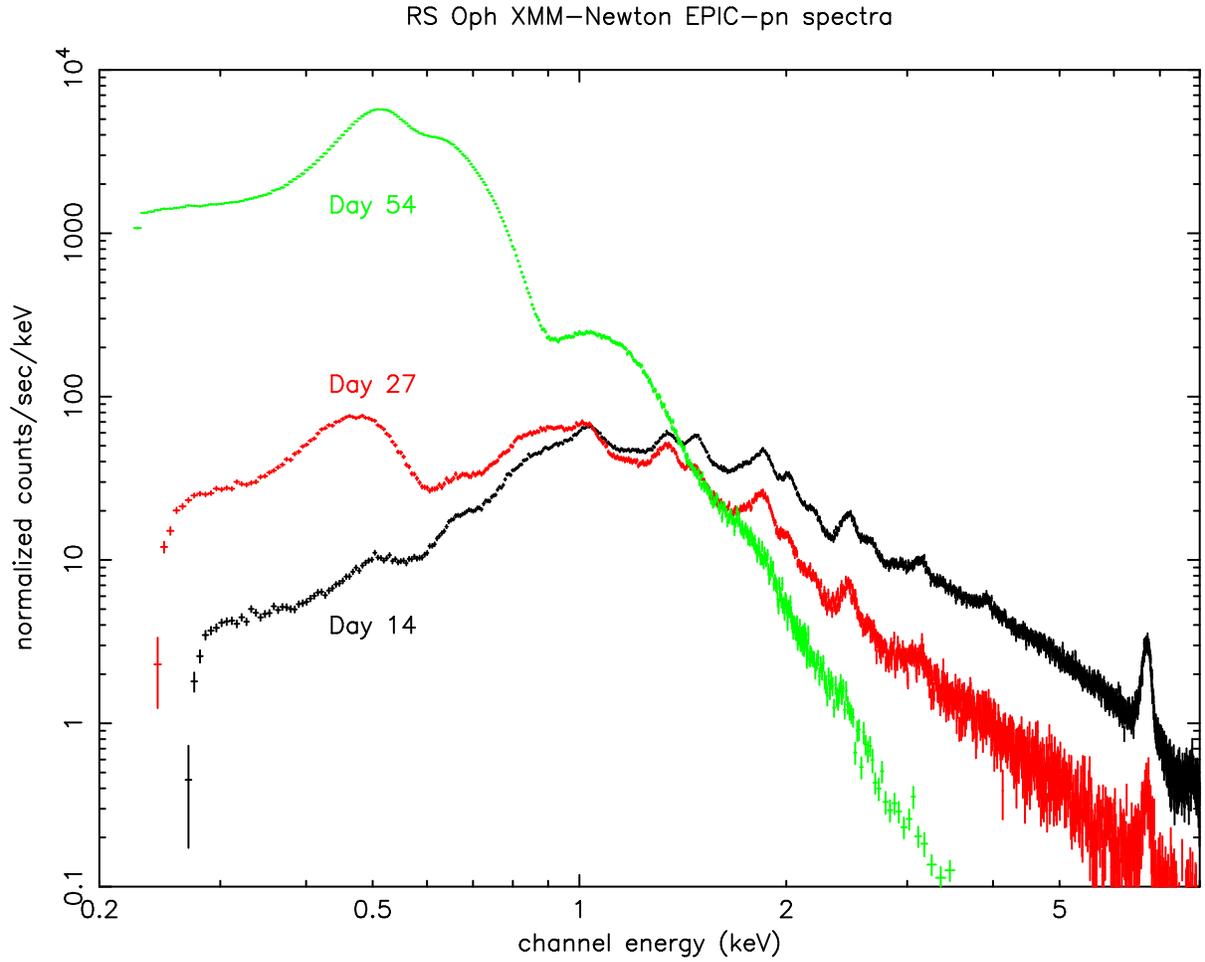}
\caption{Comparison of broadband spectra obtained with the EPIC-pn camera
onboard XMM-Newton on February 26th, March 10th and April 7th.  Note that the
count rate scale is logarithmic.  The huge increase in soft flux as the white dwarf
atmospheric emission emerges is very clear.}
\end{center}
\end{figure}
\clearpage

\begin{figure}
\begin{center}
\includegraphics{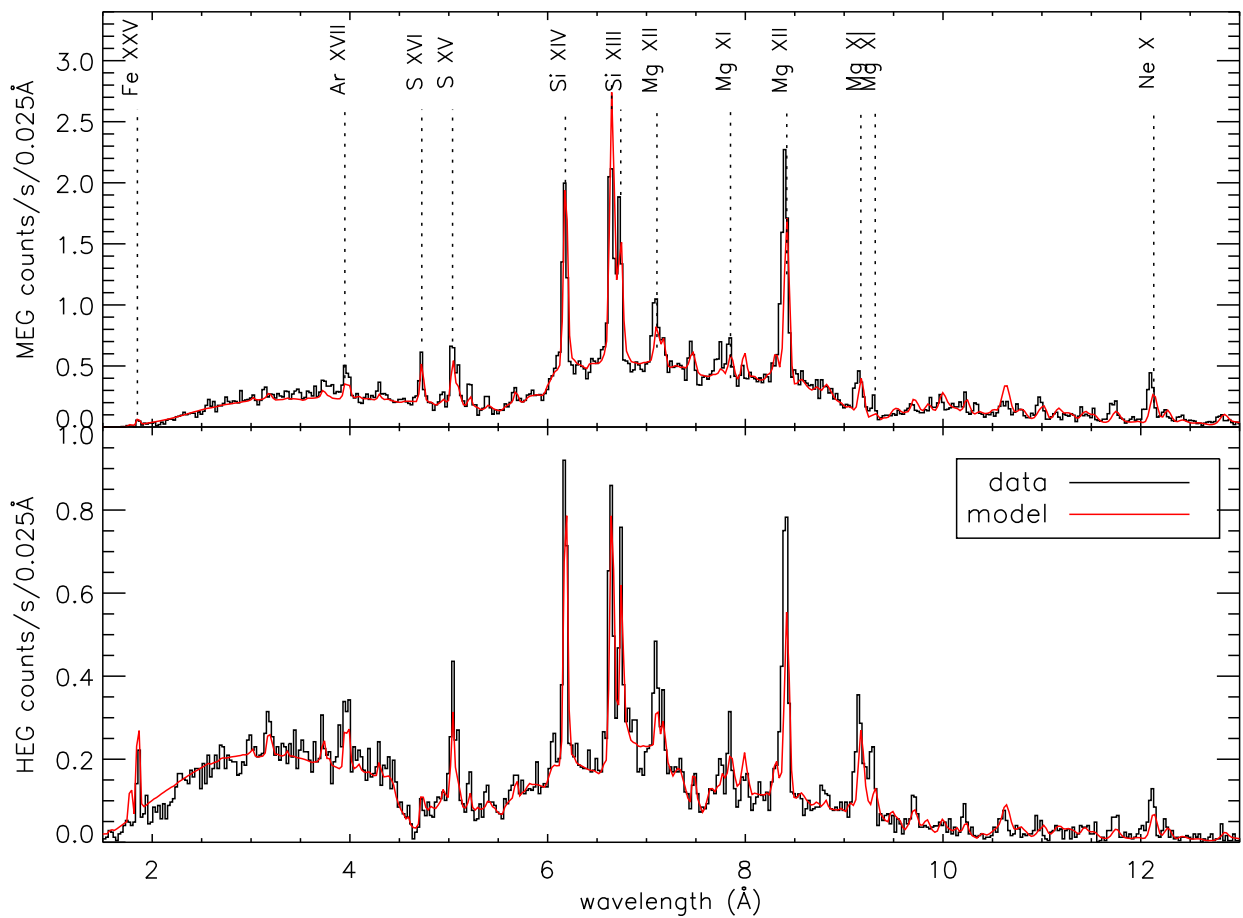}
\caption{Grating data obtained with the HETG on Chandra on February 26th, with best fit multi-temperature APEC model.  The MEG data is shown in black in the upper panel, and the HEG data (again in black) in the lower panel.  Note the different y scales on each panel.  The model, shown in red, was fit to the +1 and -1 orders of both the HEG and MEG simultaneously, although we show only the positive orders here.  The model parameters are discussed in the body of the paper.}
\end{center}
\end{figure}
\clearpage

\begin{figure}
\begin{center}
\includegraphics{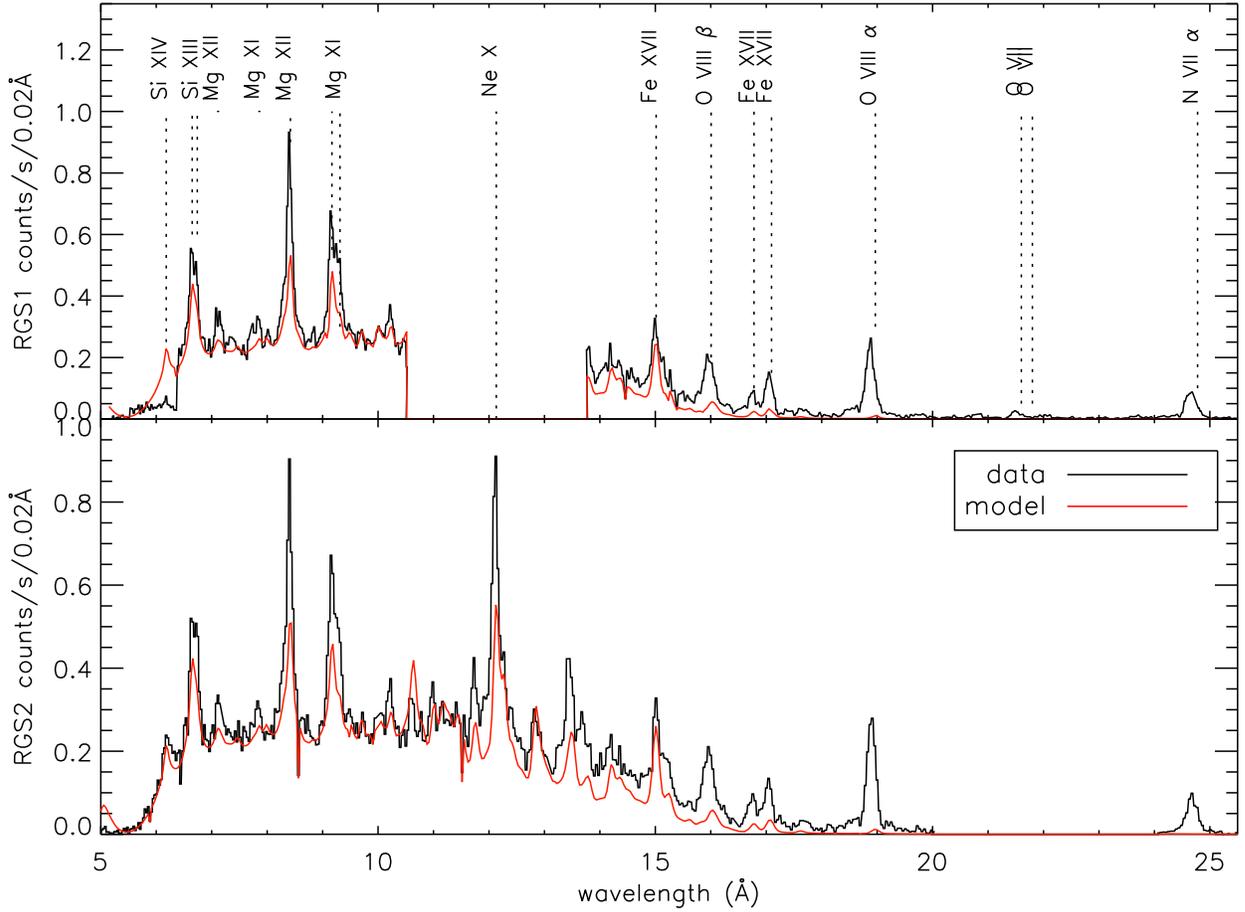}
\caption{XMM-Newton data obtained with RGS on February 26th (black).  RGS1 data is shown in the upper panel, and RGS2 in the lower panel.  Again, note the differing y scale ranges in each panel.  The multi-temperature APEC model used for the Chandra data is clearly a much poorer fit to the RGS data.  In particular, oxygen and nitrogen features are poorly accounted for by the model.  An additional low temperature component is inconsistent with the observed H-like to He-like line ratios.  The enhanced O and N features are consistent with X-ray emission from the nova ejecta.}
\end{center}
\end{figure}
\clearpage

\begin{figure}
\begin{center}
\includegraphics{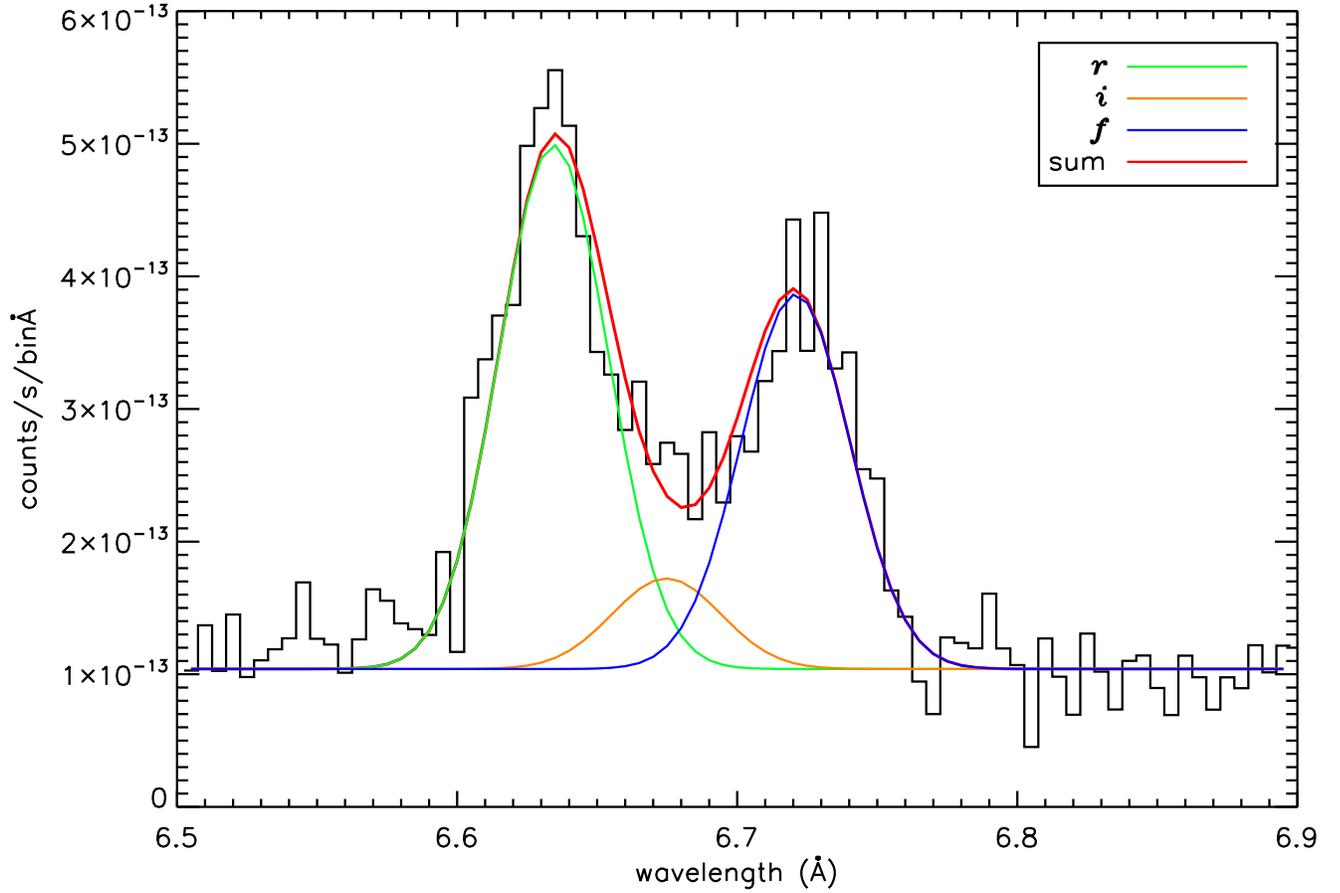}
\caption{Three gaussian component fit the the Si XIII He-like triplet in the MEG data.  
The wavelengths and widths of the intercombination (orange) and forbidden (blue) lines were constrained to those of the resonance line (green), which were then fit to the data.  The sum of the three components is given by the red line.}
\end{center}
\end{figure}
\clearpage

\begin{figure}
\begin{center}
\includegraphics[width=3in]{f5a.ps}\\
\includegraphics[width=3in]{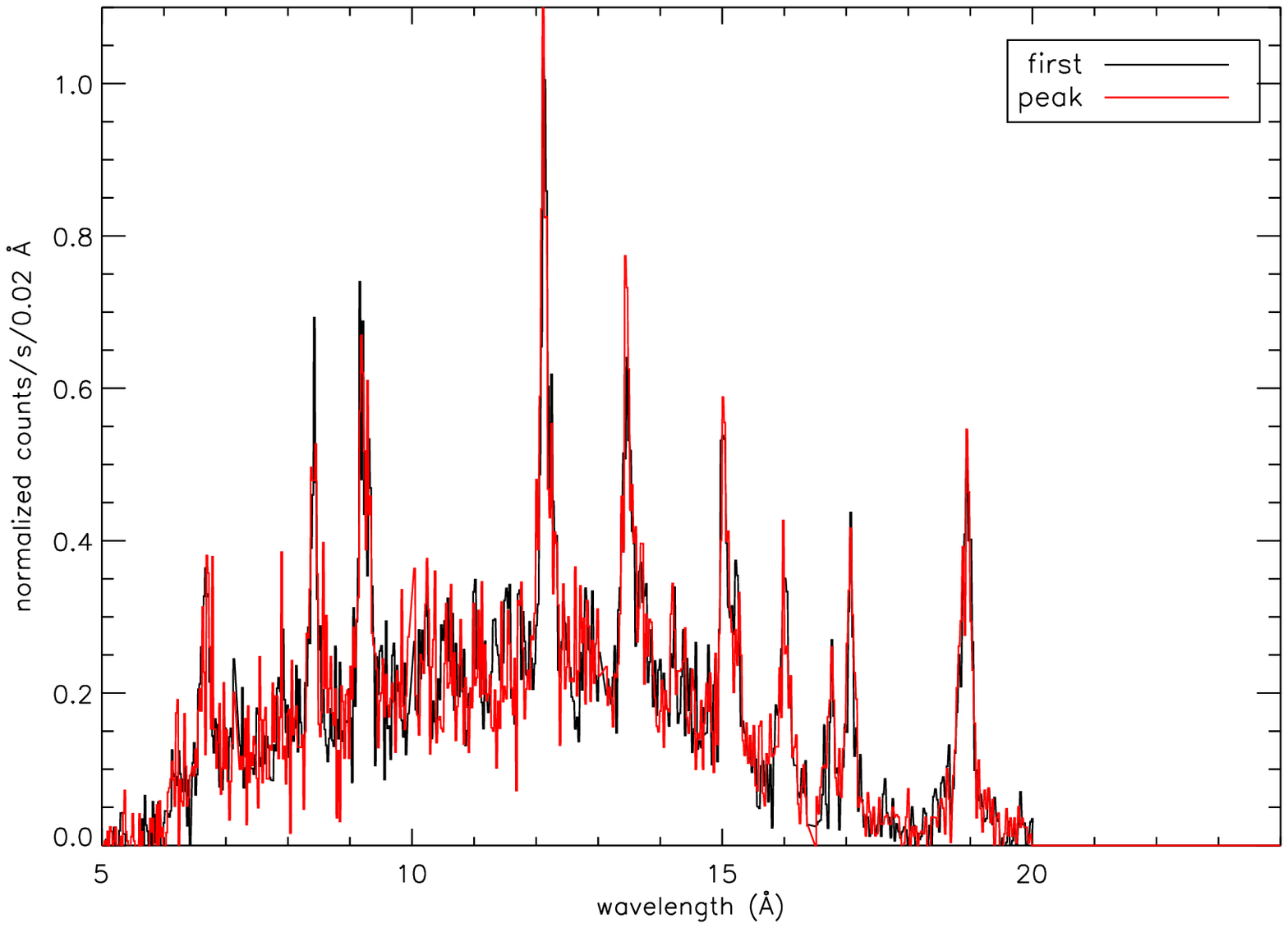}\\
\includegraphics[width=3in]{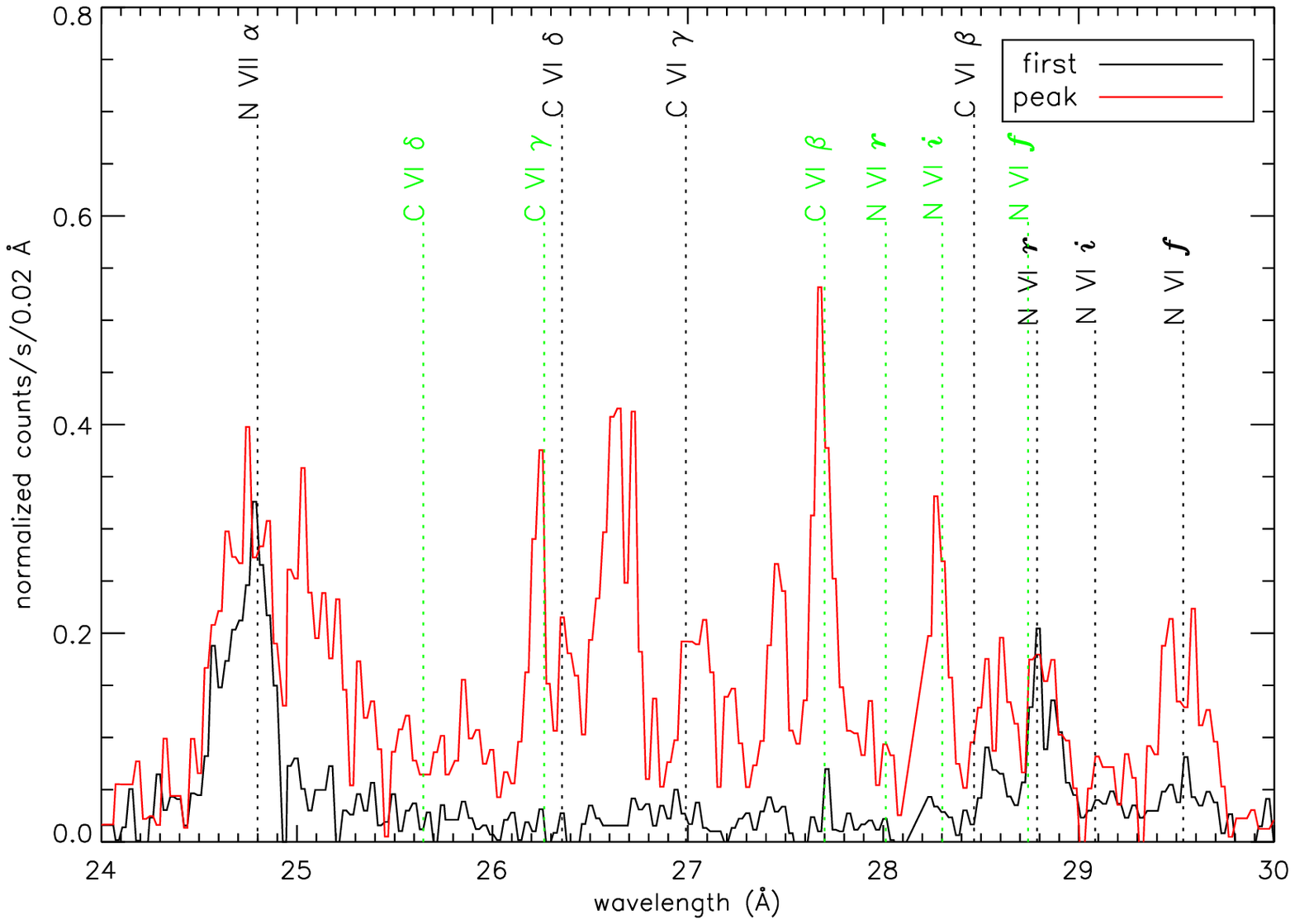} \\

\caption{Three details of the day 27 observation.  In the upper panel, we plot the EPIC-pn lightcurve in the range 0.15--0.4 keV showing the flare approximately 5000s into the observation.  In the middle and lower panels, we plot the RGS2 data comparing the spectrum before and after the flare (corresponding to time periods 1 and 3 in the text). Although we do not show them here, the RGS1 data show the same behavior.  The middle panel shows the spectrum from 5 to 24 \AA, while the bottom panel shows the 24--30 \AA\ data.  It is clear from comparing the two that no spectral changes occur shortward of 20 \AA.  In the lower panel, we give rest frame wavelength IDs in black, and IDs for the new post flare lines in green.}
\end{center}
\end{figure}
\clearpage

\begin{figure}
\begin{center}
\includegraphics[width=5in]{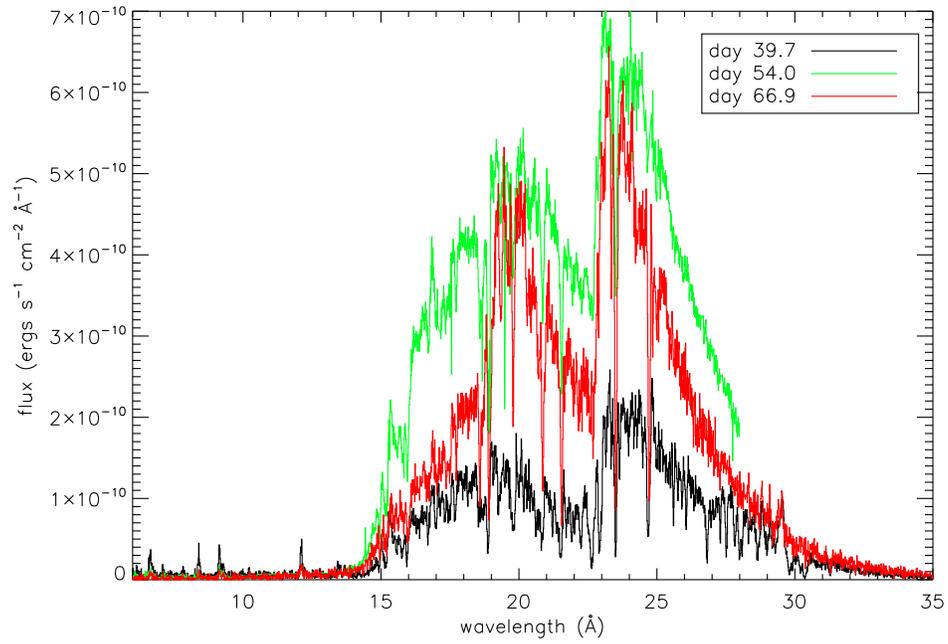} \\
\caption{Fluxed spectra of the three observations of the supersoft phase, obtained with Chandra on days 39.7 and 66.9, and with XMM-Newton on day 54.0.}
\end{center}
\end{figure}
\clearpage

\begin{figure}
\begin{center}
\includegraphics[width=5in]{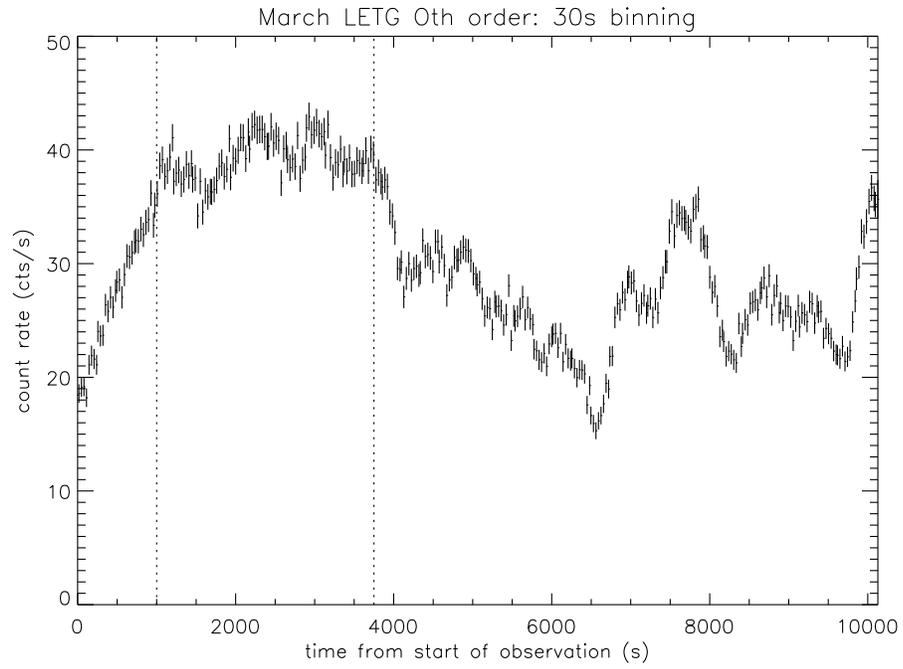} \\
\caption{0th order lightcurve for March 24th LETG observation, 30s bins.}
\end{center}
\end{figure}
\clearpage

\begin{figure}
\begin{center}
\includegraphics{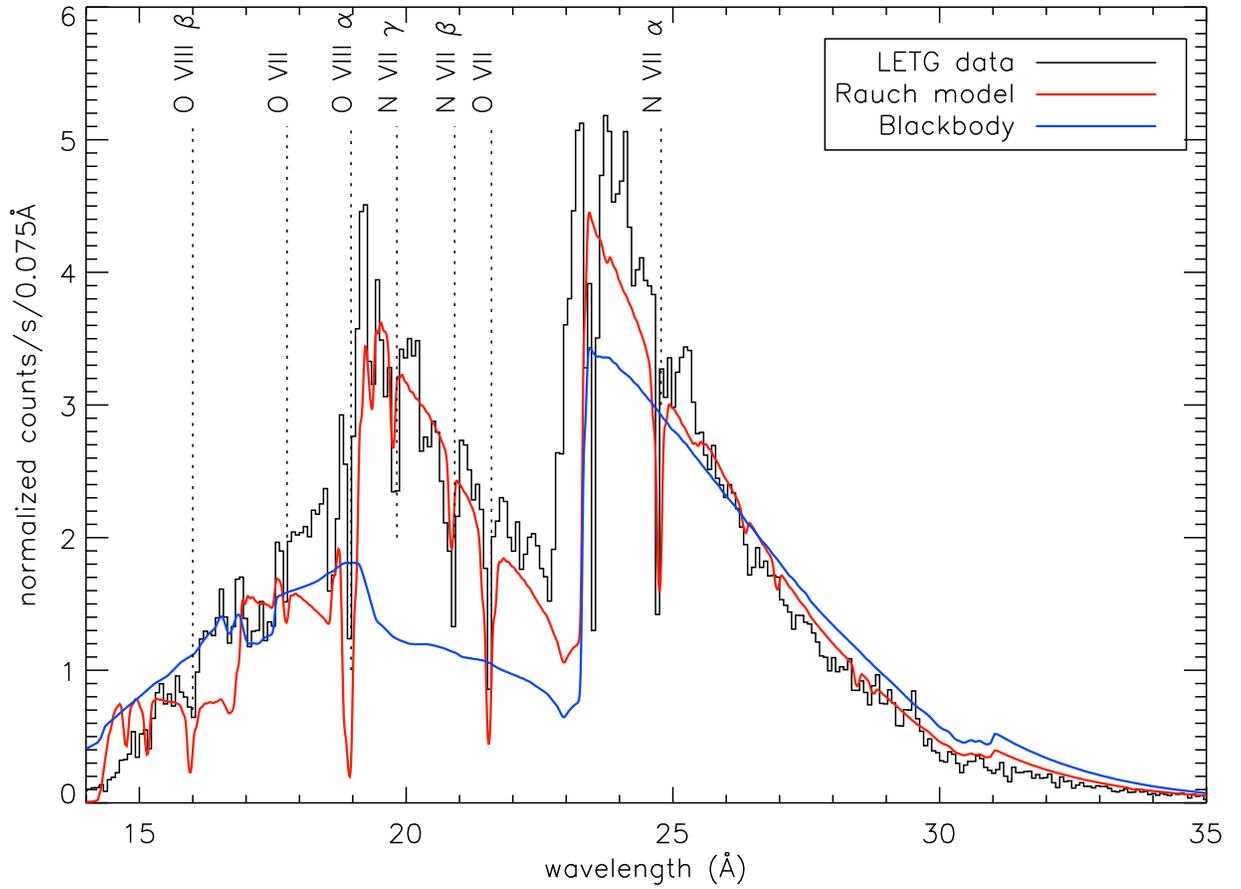}
\caption{LETG data of April 20th, with model fits.  The Rauch 
model (red) is our best fit NLTE atmospheric model, with log(g) = 9 and T = 804,000 K.  The 
model has a C/N ratio 1/1000th that of the solar value, as found in CNO cycle ashes.  The best blackbody fit (blue) is included to show the inadequacies of this simple model in fitting hot white dwarf spectra. }
\end{center}
\end{figure}
\clearpage

\begin{figure}
\begin{center}
\includegraphics{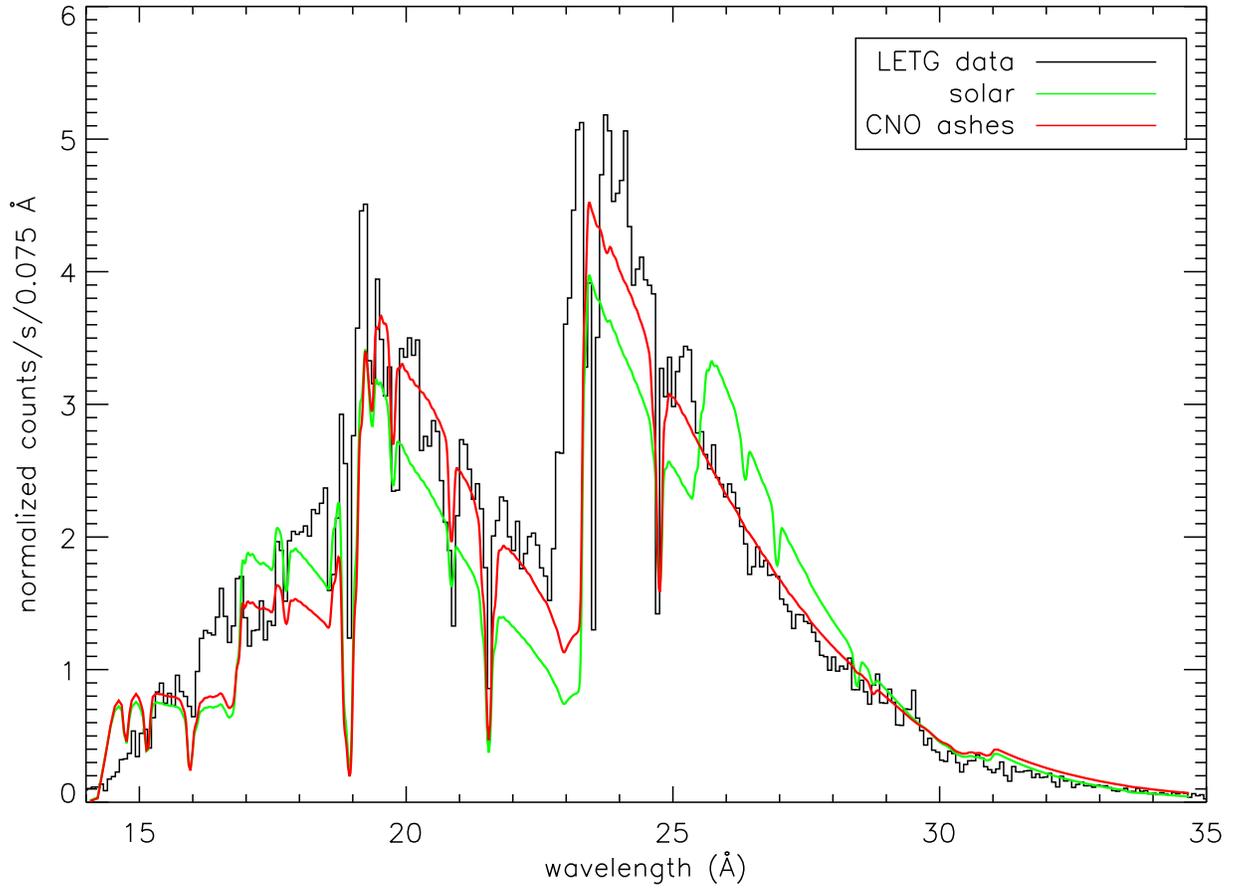}
\caption{Comparison of best fit depleted C/N model (red) with a model of solar abundance (green), showing the ability of the model to probe the surface abundances of the white dwarf.  The fit is much better with the depleted C/N model.}
\end{center}
\end{figure}
\clearpage

\begin{figure}
\begin{center}
\includegraphics[width=4in]{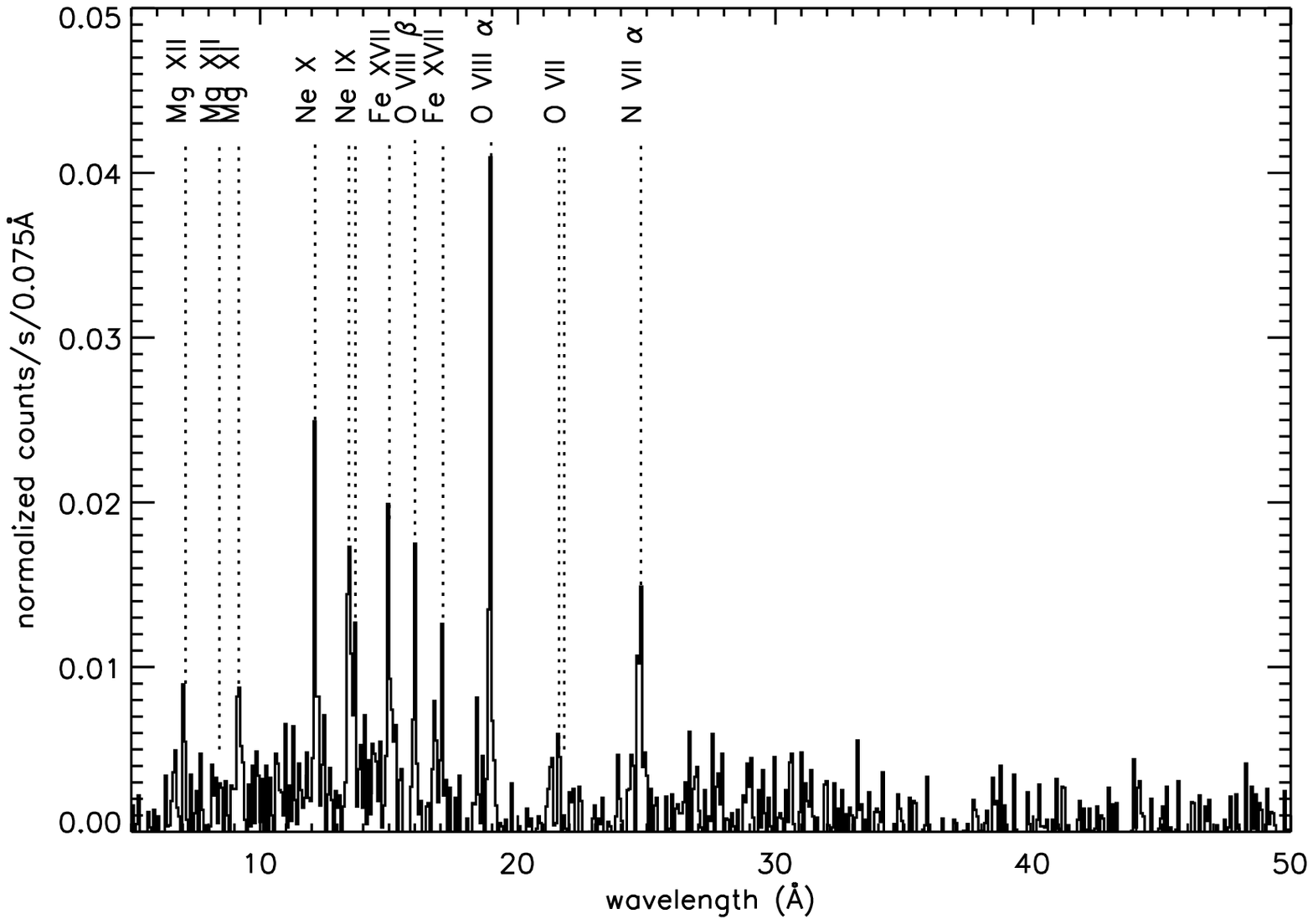}
\includegraphics[width=4in]{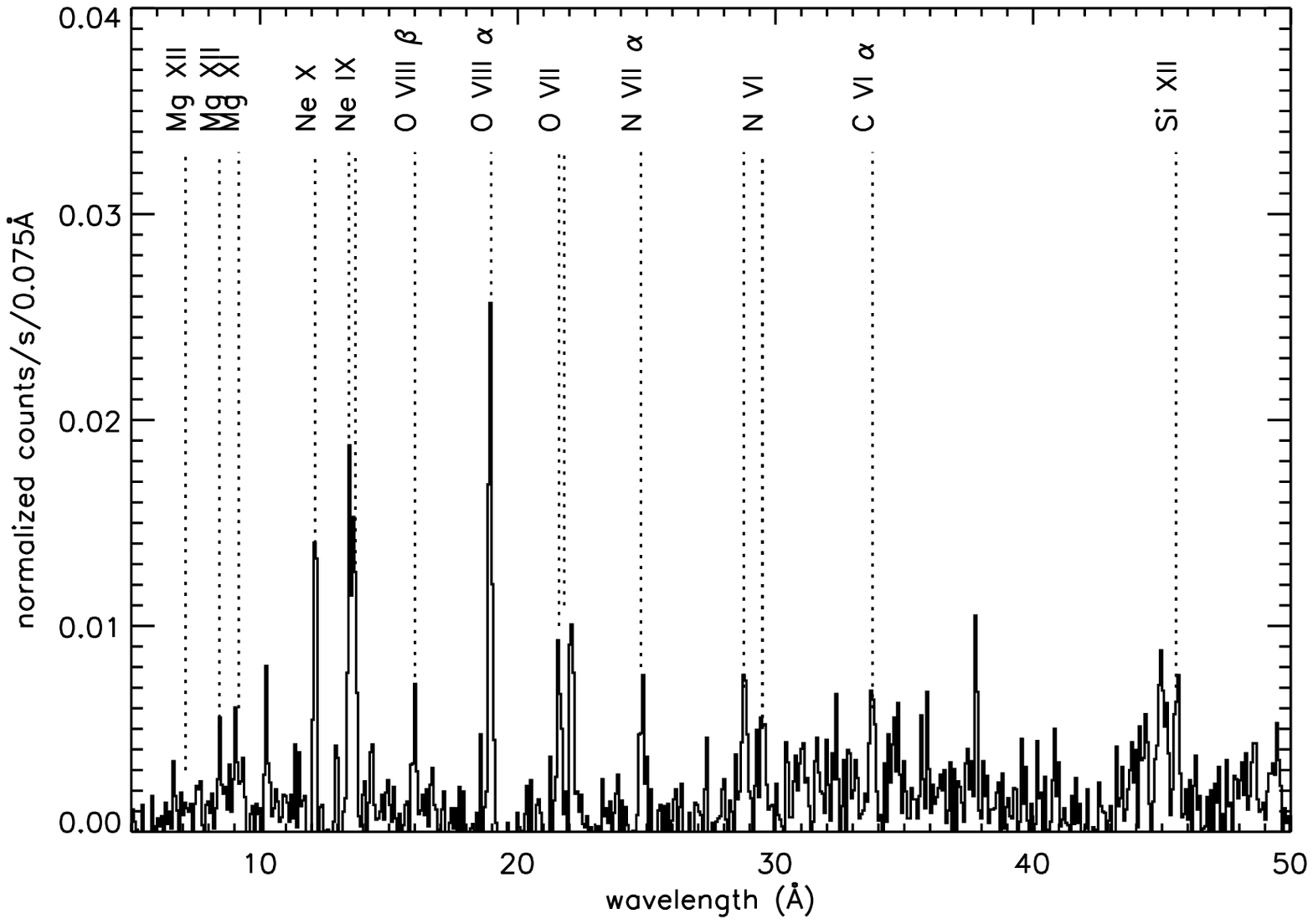}
\caption{Data obtained with the LETG on June 4th (upper panel).  Included for comparison is LETG data obtained for nova V382 Vel (lower panel).  Both spectra were obtained after the supersoft phase had ended. The identifications of the brightest lines are included.}
\end{center}
\end{figure}
\clearpage

\begin{figure}
\begin{center}
\includegraphics[width=4in]{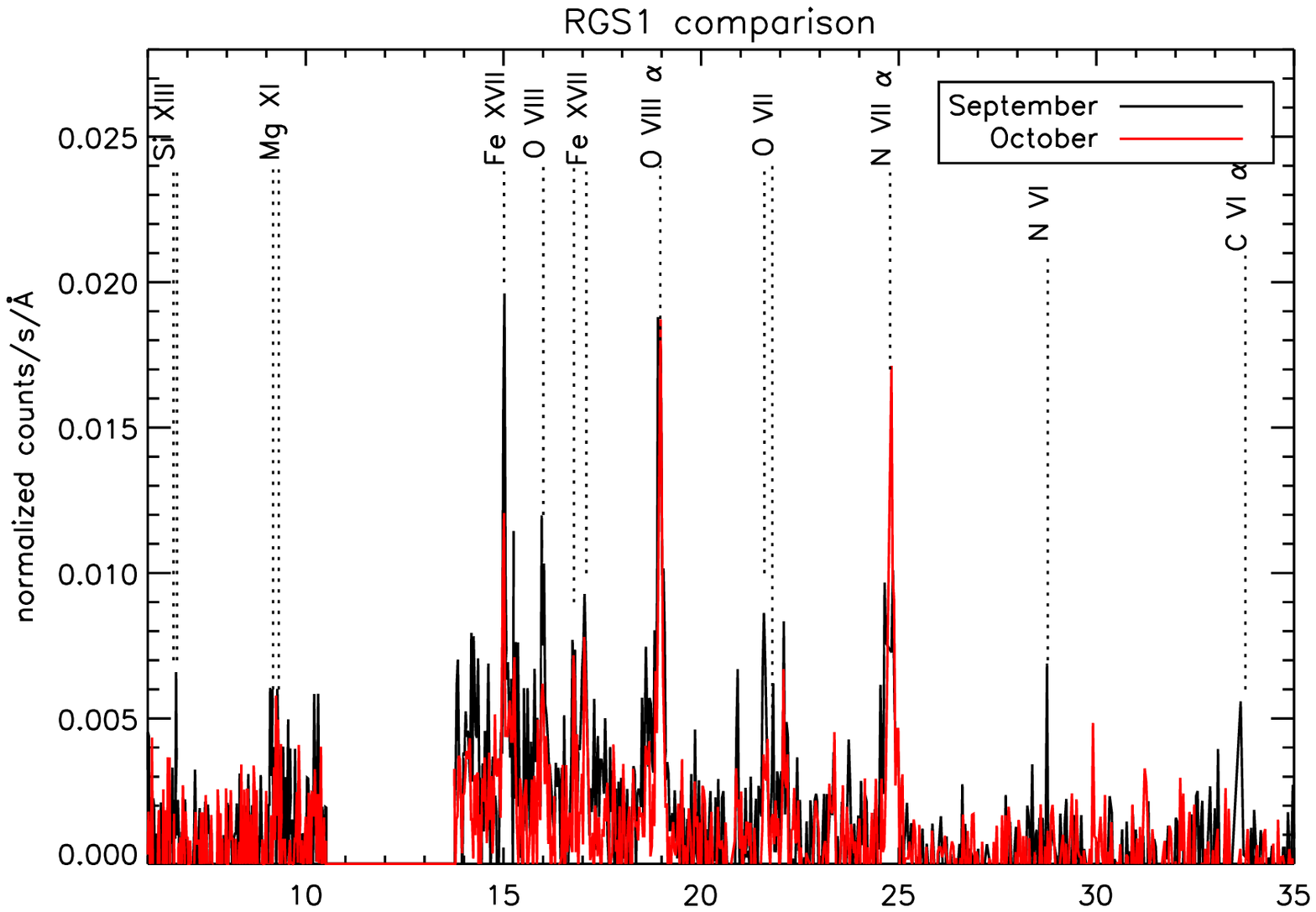}
\includegraphics[width=4in]{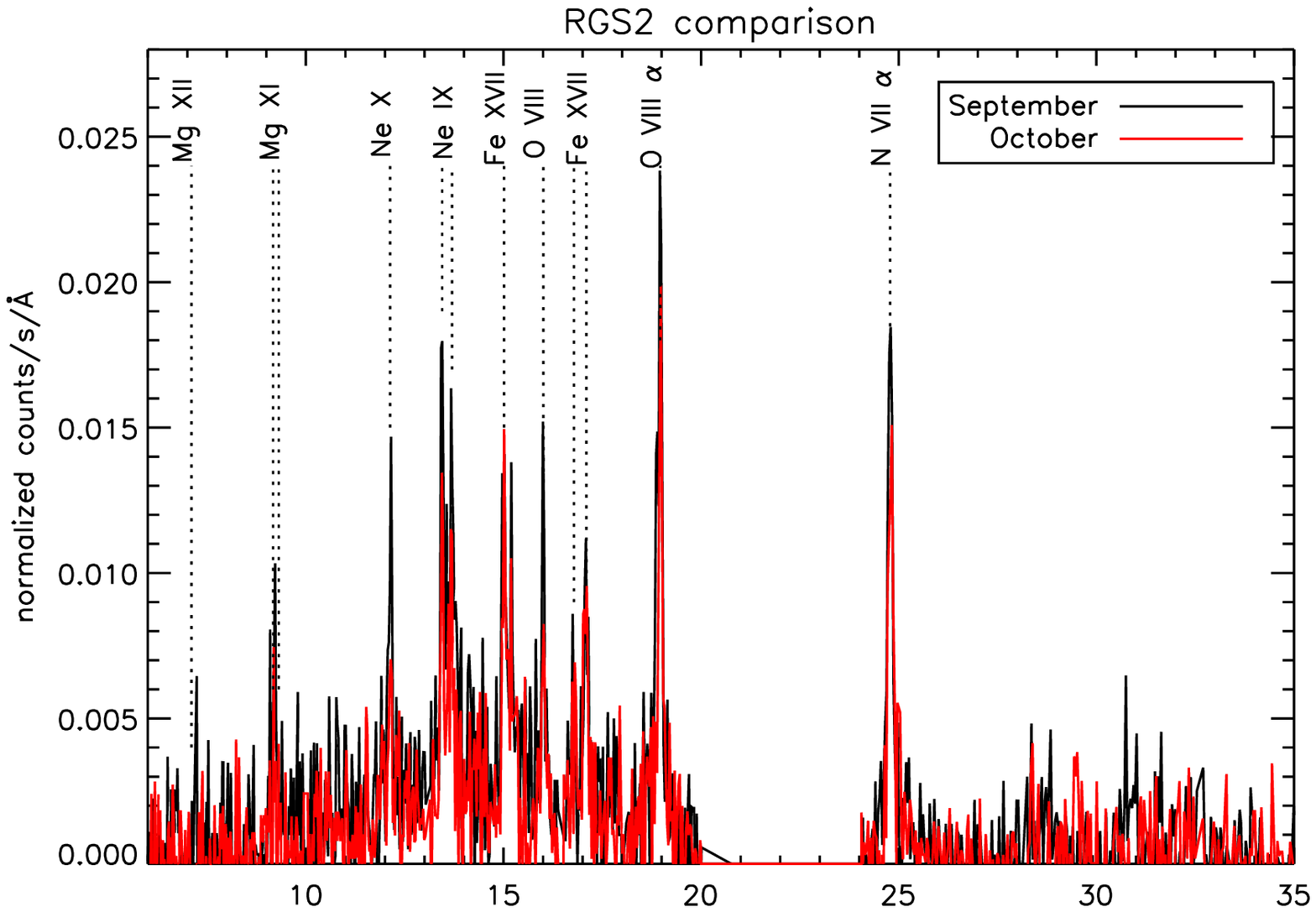}
\caption{Comparison of RGS spectra from September and October.  The upper panel shows the RGS1 data, the lower panel shows RGS2.}
\end{center}
\end{figure}
\clearpage

\begin{figure}
\begin{center}
\includegraphics[width=5in,angle=270]{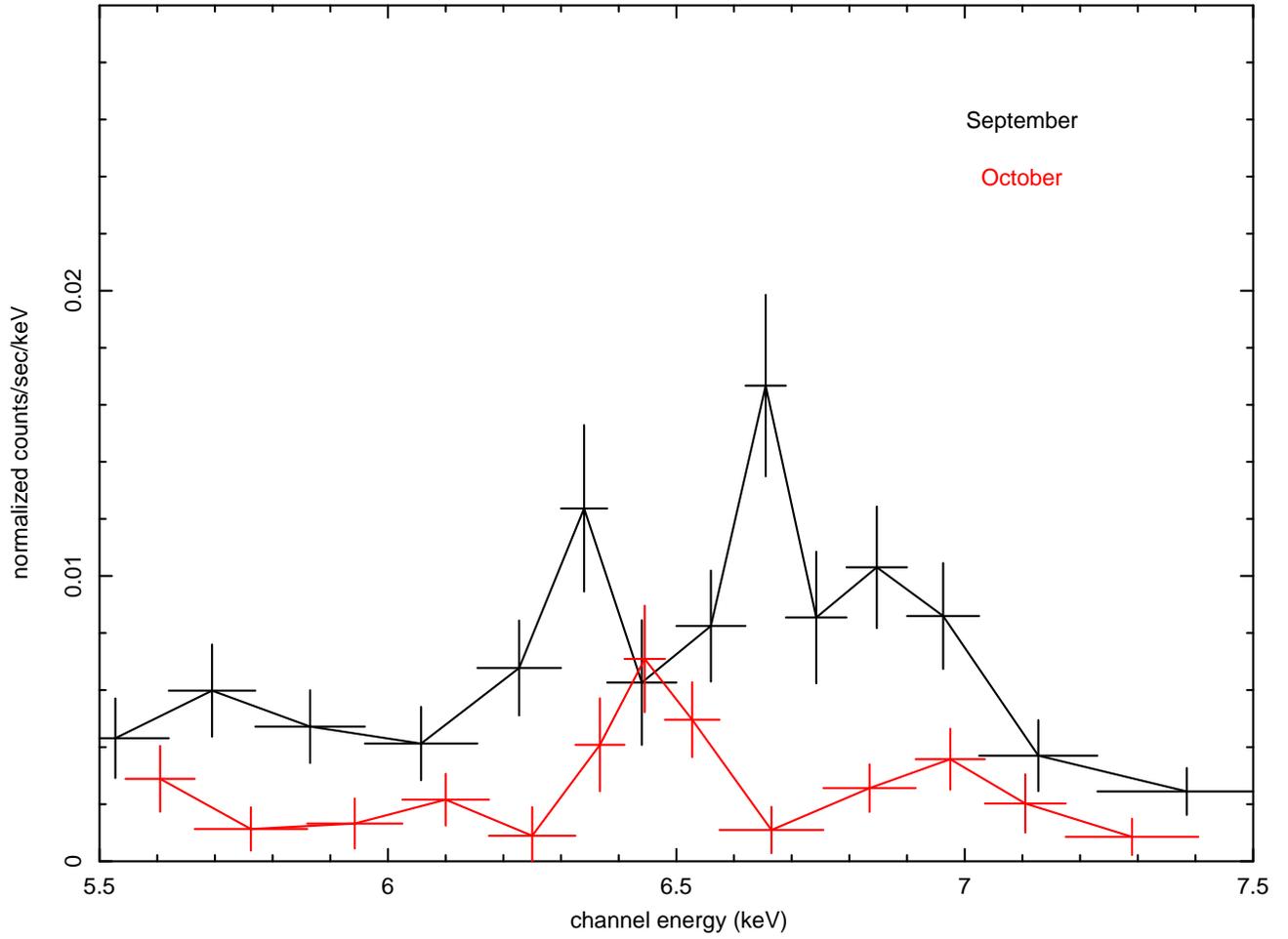}
\caption{Detail of the September (black) and October (red) EPIC-pn spectra in the energy range 5.5--7.5 keV.  Both the Fe K $\alpha$ complex at 6.4 keV, and the Fe XXV resonance line at 6.7 keV are detected in September. }
\end{center}
\end{figure}
\clearpage








\begin{thebibliography}{}

\bibitem[Althaus et al.(2005)]{2005A&A...441..689A} Althaus, L.~G., 
Garc{\'{\i}}a-Berro, E., Isern, J., \& C{\'o}rsico, A.~H.\ 2005, \aap, 441, 
689 

\bibitem[Arnaud \& Rothenflug(1985)]{1985A&AS...60..425A} Arnaud, M., \& 
Rothenflug, R.\ 1985, \aaps, 60, 425

\bibitem[Bode(1987)]{1987rorn.conf.....B} Bode, M.~F.\ 1987, in RS  Oph 
(1985) and the Recurrent Nova Phenomenon, ed. M. F. Bode (Utrect: VNU Science Press), 1

\bibitem[Bode et al.(2006)]{2006ApJ...652..629B} Bode, M.~F., et al.\ 2006, 
\apj, 652, 629 

\bibitem[Cassinelli et al.(2001)]{2001ApJ...554L..55C} Cassinelli, J.~P., 
Miller, N.~A., Waldron, W.~L., MacFarlane, J.~J., \& Cohen, D.~H.\ 2001, 
\apjl, 554, L55 

\bibitem[Cassinelli et al.(2007)]{2007_01} Cassinelli, J. et al.\ 2007, in Clumping in Hot Star Winds, editors W.-R. Hamann, A. Feldmeier \& L. Oskinova, Potsdam: Univ.-Verl.

\bibitem[Crowley(2006)]{2006PhDT.........4C} Crowley, C.\ 2006, 
Ph.D.~Thesis

\bibitem[Dobrzycka et al.(1996)]{1996AJ....111.2090D} Dobrzycka, D., 
Kenyon, S.~J., Proga, D., Mikolajewska, J., \& Wade, R.~A.\ 1996, \aj, 111, 
2090 

\bibitem[Fekel et al.(2000)]{2000AJ....119.1375F} Fekel, F.~C., Joyce, 
R.~R., Hinkle, K.~H., \& Skrutskie, M.~F.\ 2000, \aj, 119, 1375

\bibitem[Fujimoto(1982)]{1982ApJ...257..752F} Fujimoto, M.~Y.\ 1982, \apj, 
257, 752 

\bibitem[Hachisu \& Kato(2001)]{2001ApJ...558..323H} Hachisu, I., \& Kato, 
M.\ 2001, \apj, 558, 323

\bibitem[Hachisu et al.(2006b)]{2006ApJ...651L.141H} Hachisu, I., et al.\ 
2006, \apjl, 651, L141

\bibitem[Hartmann \& Heise(1997)]{1997A&A...322..591H} Hartmann, H.~W., \& 
Heise, J.\ 1997, \aap, 322, 591

\bibitem[Hjellming et al.(1986)]{1986ApJ...305L..71H} Hjellming, R.~M., van 
Gorkom, J.~H., Seaquist, E.~R., Taylor, A.~R., Padin, S., Davis, R.~J., \& 
Bode, M.~F.\ 1986, \apjl, 305, L71 

\bibitem[Houck \& Denicola(2000)]{2000ASPC..216..591H} Houck, J.~C., \& 
Denicola, L.~A.\ 2000, Astronomical Data Analysis Software and Systems IX, 
216, 591 

\bibitem[Leibowitz et al.(2006)]{2006MNRAS.371..424L} Leibowitz, E., Orio, 
M., Gonzalez-Riestra, R., Lipkin, Y., Ness, J.-U., Starrfield, S., Still, 
M., \& Tepedelenlioglu, E.\ 2006, \mnras, 371, 424 

\bibitem[Leibowitz et al.(2007)]{2007_02} Leibowitz, E. et al.\ 2007, submitted to \apj

\bibitem[Mason et al.(1987)]{1987rorn.conf..167M} Mason, K.~O., 
C{\'o}rdova, F.~A., Bode, M.~F., \& Barr, P.\ 1987, in RS  Oph (1985) and 
the Recurrent Nova Phenomenon, ed. M. F. Bode (Utrecht: VNU Science Press), 167 

\bibitem[Monnier et al.(2006)]{2006ApJ...647L.127M} Monnier, J.~D., et al.\ 
2006, \apjl, 647, L127

\bibitem[Mukai et al.(2003)]{2003ApJ...586L..77M} Mukai, K., Kinkhabwala, 
A., Peterson, J.~R., Kahn, S.~M., \& Paerels, F.\ 2003, \apjl, 586, L77

\bibitem[M{\"u}rset \& Schmid(1999)]{1999A&AS..137..473M} M{\"u}rset, U., 
\& Schmid, H.~M.\ 1999, \aaps, 137, 473 

\bibitem[Narumi et al.(2006)]{2006IAUC.8671....2N} Narumi, H., Hirosawa, 
K., Kanai, K., Renz, W., Pereira, A., Nakano, S., Nakamura, Y., \& 
Pojmanski, G.\ 2006, \iaucirc, 8671, 2 

\bibitem[Ness et al.(2003)]{2003ApJ...594L.127N} Ness, J.-U., et al.\ 2003, 
\apjl, 594, L127 

\bibitem[Ness et al.(2005)]{2005MNRAS.364.1015N} Ness, J.-U., Starrfield, 
S., Jordan, C., Krautter, J., \& Schmitt, J.~H.~M.~M.\ 2005, \mnras, 364, 
1015 

\bibitem[Ness et al.(2007)]{2007arXiv0705.1206N} Ness, J.-U., et al.\ 2007, 
\apj, in press

\bibitem[Oppenheimer \& Mattei(1993)]{1993AAS...183.5503O} Oppenheimer, B., 
\& Mattei, J.~A.\ 1993, Bulletin of the American Astronomical Society, 25, 
1378 

\bibitem[Oppenheimer \& Mattei(1996)]{1996IAUS..165..457O} Oppenheimer, 
B.~D., \& Mattei, J.~A.\ 1996, IAU Symp.~165: Compact Stars in Binaries, 
165, 457

\bibitem[Orio et al.(2001)]{2001MNRAS.326L..13O} Orio, M., et al.\ 2001, 
\mnras, 326, L13

\bibitem[Orio et al.(2001)]{2001A&A...373..542O} Orio, M., Covington, J. \&
\"{O}gelman, H.\ 2001, \aap, 373, 542

\bibitem[Osborne et al.(2006a)]{2006ATel..764....1O} Osborne, J., et al.\ 
2006, The Astronomer's Telegram, 764, 1 

\bibitem[Osborne et al.(2006b)]{2006ATel..770....1O} Osborne, J., et al.\ 
2006, The Astronomer's Telegram, 770, 1 

\bibitem[Osborne et al.(2006c)]{2006ATel..801....1O} Osborne, J., et al.\ 
2006, The Astronomer's Telegram, 801, 1 

\bibitem[Osborne et al.(2006d)]{2006ATel..838....1O} Osborne, J., et al.\ 
2006, The Astronomer's Telegram, 838, 1 

\bibitem[Paczy{\'n}ski(1971)]{1971AcA....21..417P} Paczy{\'n}ski, B.\ 1971, 
Acta Astronomica, 21, 417 

\bibitem[Porquet et al.(2001)]{2001A&A...376.1113P} Porquet, D., Mewe, R., 
Dubau, J., Raassen, A.~J.~J., \& Kaastra, J.~S.\ 2001, \aap, 376, 1113 

\bibitem[Prialnik et al.(1978)]{1978A&A....62..339P} Prialnik, D., Shara, 
M.~M., \& Shaviv, G.\ 1978, \aap, 62, 339

\bibitem[Rauch(2003)]{2003A&A...403..709R} Rauch, T.\ 2003, \aap, 403, 709

\bibitem[Rauch et al.(2005)]{2005AIPC..774..361R} Rauch, T., Werner, K., \& 
Orio, M.\ 2005, AIP Conf.~Proc.~774: X-ray Diagnostics of Astrophysical 
Plasmas: Theory, Experiment, and Observation, 774, 361 

\bibitem[Rauch et al.(2005)]{2005ASPC..334..423R} Rauch, T., Orio, M., 
Gonzales-Riestra, R., \& Still, M.\ 2005, ASP Conf.~Ser.~334: 14th European 
Workshop on White Dwarfs, 334, 423 

\bibitem[Schaefer(2004)]{2004IAUC.8396....2S} Schaefer, B.~E.\ 2004, 
\iaucirc, 8396, 2 

\bibitem[Schatzman(1950)]{1950AnAp...13..384S} Schatzman, E.\ 1950, Annales 
d'Astrophysique, 13, 384 

\bibitem[Sokoloski et al.(2006)]{2006Natur.442..276S} Sokoloski, J.~L., 
Luna, G.~J.~M., Mukai, K., \& Kenyon, S.~J.\ 2006, \nat, 442, 276 

\bibitem[Starrfield et al.(1974)]{1974ApJS...28..247S} Starrfield, S., 
Sparks, W.~M., \& Truran, J.~W.\ 1974, \apjs, 28, 247

\bibitem[Waldron \& Cassinelli(2007)]{2007arXiv0707.0024W} Waldron, W.~L., 
\& Cassinelli, J.~P.\ 2007, \apj, in press

\bibitem[Yao \& Wang(2006)]{2006ApJ...641..930Y} Yao, Y., \& Wang, Q.~D.\ 
2006, \apj, 641, 930 

\end{thebibliography}
\end{document}